\documentclass[onecolumn]{IEEEtran}
\usepackage{amsmath,amsfonts}
\usepackage{algorithmic}
\usepackage{algorithm}
\usepackage{array}
\usepackage[caption=false,font=normalsize,labelfont=sf,textfont=sf]{subfig}
\usepackage{textcomp}
\usepackage{stfloats}
\usepackage{url}
\usepackage{verbatim}
\usepackage{graphicx}
\usepackage{cite}
\usepackage{graphicx,amsthm,amssymb,amsmath,authblk}
\usepackage{hyperref,yhmath,cleveref}
\usepackage{multirow}

\crefname{figure}{Fig.}{Figs.}
\crefname{definition}{Def.}{Defs.}
\crefname{equation}{Eq.}{Eqs.}
\crefname{algorithm}{Alg.}{Algs.}
\crefname{lemma}{Lemma}{Lemmas}
\crefname{theorem}{Theorem}{Theorems}
\crefname{section}{Section}{Sections}
\crefname{proposition}{Proposition}{Propositions}

\newtheorem{theorem}{Theorem}
\newtheorem{lemma}[theorem]{Lemma}

\title{Practical implementation of geometric quasi-cyclic LDPC codes}
\author{Simeon~Ball and Tomàs~Ortega %
\thanks{This research is supported by the Spanish Ministry of Science, Innovation and Universities grant PID2020-113082GB-I00 funded by MICIU/AEI/10.13039/501100011033.}%
\thanks{Simeon~Ball is with the Department of Mathematics, Universitat Politecnica Catalunya, 08034 Barcelona (email:simeon.michael.ball@upc.edu).}%
\thanks{Tomas~Ortega is with the Department of Electrical Engineering and Computer Science, University of California, Irvine, CA 92697 USA (email: tomaso@uci.edu).}
}
\begin{document}

\maketitle

\begin{abstract}
	We detail for the first time a complete explicit description of the quasi-cyclic structure of all classical finite generalized quadrangles. Using these descriptions we
	construct families of quasi-cyclic LDPC codes derived from the point-line incidence matrix of the quadrangles by explicitly calculating quasi-cyclic generator and parity check matrices for these codes. This allows us to construct parity check and generator matrices of all such codes of length up to 400000. These codes cover a wide range of transmission rates, are easy and fast to implement and perform close to Shannon's limit with no visible error floors. We also include some performance data for these codes. Furthermore, we include a complete explicit description of the quasi-cyclic structure of the point-line and point-hyperplane incidences of the finite projective and affine spaces.
\end{abstract}

\section{Introduction}
In many modern communication systems Low Density Parity Check (LDPC) codes are used. LDPC codes are those codes for which the number of ones in the check matrix is very small compared to the size of the matrix. The low density nature of the check matrix allows one to implement fast decoding algorithms, such as belief propagation and sum-product algorithms. To be able to reach performance levels comparable to Shannon's limit, quasi-cyclic check matrices are used.
% This allows one to implement long LDPC codes, in other words the length $n$ of the code is large, typically between 2000 and 20000 bits.
This allows one to implement long LDPC codes, in other words the length $n$ of the code is large -- over 64,000 bits long in standard satellite broadcast communications, for example.
A quasi-cyclic LDPC $m \times n$ check matrix H can be described by a block size $b$ and a $(m/b) \times (n/b)$ matrix H$^{\mathrm{rep}}$, whose entries are subsets $\mathrm{H}_{ij}$ of $\{0,\ldots,b-1\}$, where $i \in \{1,\ldots,(m/b)\}$ and $j \in \{1,\ldots,(n/b)\}$. Typically, the subset $\mathrm{H}_{ij}$ is empty which corresponds to the $b\times b$ zero matrix in H in the $(i,j)$ cell. A singleton subset $\mathrm{H}_{ij}=\{r\}$ indicates that in the $(i,j)$ cell we have a copy of the $b \times b$ identity matrix shifted $r$ bits (cyclically) to the right. A larger subset will involve a superposition of such shifts of the identity matrix.
% This representation of the quasi-cyclic LDPC check matrix H allows one to implement the decoding algorithms mentioned above efficiently. 
This representation of the quasi-cyclic LDPC check matrix H allows one to implement decoding algorithms, such as the Sum-Product Algorithm (SPA), efficiently.

The Tanner graph $\Gamma$ is the bipartite graph with stable sets of size $m$ and $n$, whose edges correspond to a one entry in the matrix H. The decoding algorithms mentioned in the previous paragraph work well if the girth of $\Gamma$, the length of the shortest cycle, is large, and decode quickly if $\Gamma$ has low diameter~\cite{kou2001low,LP2005}. The diameter is the maximum distance between any two vertices. These conflicting objectives are optimized when the girth is twice the diameter. Graphs achieving this bound are incidence matrices of generalized polygons. The rows of $\mathrm{H}$ are indexed by the points of the polygon and the columns are indexed by the lines, or vice-versa. Finite generalized polygons have diameter $3$, $4$, $6$ or $8$, see \cite{FH1964}, and are respectively called, projective planes, generalized quadrangles, generalized hexagons and generalized octagons. The LDPC code used in IEEE 802.3 standard (2048,1723) LDPC code for the 10-G Base-T Ethernet, is a quasi-cyclic LDPC code from an affine plane over the field with 32 elements.
The code has block size $b=64$, length $n=2048$ and dimension $k=1723$, see \cite[Example 10.5]{RL2009}.
The LDPC code used in the NASA Landsat Data Continuation is a quasi-cyclic LDPC code from a $3$-dimensional affine space which has block size $b=511$, length $n=8176=16b$ and dimension $k=7154=14b$, see \cite[Example 10.10]{RL2009}.

The rest of the paper is organized as follows:
In \cref{sec:qc-generator-and-check} we will present the basic structure of quasi-cyclic generator and check matrices we will use throughout this work.
In \cref{ellipticexample}, we will describe how to efficiently employ quasi-cyclic LDPC codes derived from the quasi-cyclic structure of the classical generalized quadrangles, which we calculate in \cref{sec:gq}. These codes are fast and efficient, can be extremely long, and perform favorably compared to commercially used codes with similar parameters. In \cref{sectionpanda}, we give the quasi-cyclic structure of the point-line and point-hyperplane incidences of the finite projective and affine spaces. In \cref{sectionparam} we list the parameters of all codes derived from classical generalized quadrangles of length up to 400000. In \cref{sectionperform} we give some performance data for particular codes in these families. Finally, in \cref{sectionfurther}, we list some possible future work that may be fruitful in this direction.

\section{Quasi cyclic generator and check matrices} \label{sec:qc-generator-and-check}

It was proven in \cite{LP2005} that the classical generalized quadrangles, of which there are six types, have a quasi-cyclic representation. However, up until now, no description of these quasi-cyclic representations was known. In this article we give a simple, explicit description of a quasi-cyclic representation for all the classical generalized quadrangles. We also include quasi-cyclic representations for incidences between points and hyperplanes and points and lines in projective and affine spaces. Using this representation, we describe how to employ the corresponding quasi-cyclic LDPC code efficiently. In many cases, we do not take the entire geometry but a carefully chosen large sub-structure, which allows us to increase the size of the blocks. It is advantageous to have a large block size since this allows the implementation of significantly longer codes. The block size is increased by the removal of a {\em spread}, a set of pairwise non-intersecting lines which cover all the points. As evidenced in the proof of Shannon's theorem, the implementation of long codes brings the performance of the code close to Shannon's limit.

Once we have described how to construct the quasi-cyclic representation of the check matrix H in a purely algebraic manner, we can compute H for classical generalized quadrangles LDPC codes of length up to 400000.

Let $C$ denote the binary linear code whose check matrix is $\mathrm{H}$. We shall refer to $C$ as the {\em full code}. Efficient encoding can be implemented if one can find a generator matrix for the code in quasi-cyclic form, see \cite{LCZL2006}.  However, such a generator matrix for the full code $C$ does not generally exist, so we take a large subcode $C'$ of $C$ for which there is a generator matrix G in quasi-cyclic form. We will call $C'$ the {\em implementable code}. A $k \times n$ generator matrix is in standard form if it has $k \times k$ submatrix which is an identity matrix. For each quadrangle and each $q$, we compute a generator matrix G in standard quasi-cyclic form for the implementable code $C'$. Note that $\mathrm{H}$ is also a check matrix for $C'$. The matrix G can be described by a $(k/b) \times ((n-k)/b)$ matrix P$^{\mathrm{rep}}$, where the $(i,j)$ entry of P$^{\mathrm{rep}}$ is $\mathrm{P}_{ij}$, a vector of $\{0,1\}^b$. Replacing each first row, with the full circulant $b \times b$ matrix one obtains a matrix $\mathrm{P}$, where
\begin{equation} \label{genmatrix}
	\mathrm{G} =(\mathrm{P}\ | \ \mathrm{id}).
\end{equation}
Here, $\mathrm{id}$ denotes the $k \times k$ identity matrix.

Given the matrix P$^{\mathrm{rep}}$, a shift-register-adder-accumulator (SRAA) circuit with a $b$-bit feedback shift register can be implemented to calculate each block of $b$ parity check bits of the encoded codeword, see \cite[Figure 1]{LCZL2006}. In series, this gives an encoding circuit of $(n-k)/b$ SRAA circuits with a total of $2(n-k)$ flip-flops, $n-k$ AND gates, and $n-k$ two-input XOR gates. The encoding is completed in a time proportional to $n-k$, see \cite[Figure 2]{LCZL2006}. An encoder which completes in $n-k$ clock cycles and $k/b$ feedback shift registers, each with $b$ flip-flops can be implemented when the circuits are put in parallel, see \cite[Figure 3]{LCZL2006}.

This gives an efficient encoding and decoding of very long quasi-cyclic LDPC codes whose performance is close to Shannon's limit.

\section{Quasi-cyclic LDPC codes from classical geometries.} \label{ellipticexample}

In this section, we will detail the method by which we build the check and generator matrices for the LDPC codes we construct from a classical geometry. We will use the elliptic quadrangle, and more specifically the elliptic quadrangle with a spread removed, by means of an example.  The method is the same for the other geometries, the only difference being that equation (\ref{ellipticeqn}) should be replaced by the corresponding equation describing the quasi-cyclic structure of the geometry.

Throughout $q=p^h$ is a prime power, where $p$ is a prime and $h$ is a positive integer. The finite field with $q$ elements will be denoted by ${\mathbb F}_q$.

Let $\omega$ be a primitive element of ${\mathbb F}_{q^{6}}$, so the powers of $\omega$ generate all the non-zero elements of the field. As we shall prove in \cref{qcrepelliptic}, the block size for the elliptic quadrangle with a spread removed is $b=q^3+1$.

We define
$$
	P=\{ x \in {\mathbb F}_{q^{6}} \ | \ 1+x^{q^3+1}+x^{(q^3+1)(q+1)}=0.\}
$$
and
$$
	L=\{ a \in {\mathbb F}_{q^{6}} \ | \ 1+a^{q^3+1}+a^{q^4+q}+a^{q^5+q^2}=0\}.
$$
Let $P_1$ be a subset of $P$ of size $q+1$ such that for all $x_0,x_1 \in P_1$, $x_0^{q^3+1} \neq x_1^{q^3+1}$.
Let $L_1$ be a subset of $L$ of size $q^2$ with the property that for $a_0,a_1 \in L_1$, $a_0^{q^3+1} \neq a_1^{q^3+1}$.

The check matrix $\mathrm{H}$ is fully described by H$^{\mathrm{rep}}$, a $(q+1) \times q^2 $ matrix whose rows are indexed by elements of $P_1$ and whose columns are indexed by elements of $L_1$. For $x \in P_1$ and $a \in L_1$, the $\mathrm{H}_{a,x}$ entry
of H$^{\mathrm{rep}}$ is a subset $\mathrm{H}^{\mathrm{rep}}_{a,x}$ of $\{0,\ldots,b-1\}$ whose elements indicate a $1$-coordinate of the first row of the circulant matrix occupying the corresponding position in the quasi-cyclic matrix H.

Let $\alpha=\omega^{q^3-1}$, so that $\alpha$ is a primitive $b$-th root of unity in ${\mathbb F}_{q^{6}}$,

The quasi-cyclic representation of the elliptic quadrangle is given by $i \in \mathrm{H}^{\mathrm{rep}}_{x,a}$ if and only if
\begin{equation} \label{ellipticeqn}
	a(\alpha^{i}x)^{q+1}-\alpha^{i}x+a^{q^2}=0,
\end{equation}
where $i$ runs from $0$ to $b-1$, see \cref{qcrepelliptic}.

We will use $q=3$ as an example, so the block size is $b=28$ and $\omega$ is a primitive element of ${\mathbb F}_{3^6}$.

For $q=3$, one computes that
$$
	P_1=\{ \omega^2, \omega^{474}, \omega^{616}, \omega^{572}\}
$$
and that
$$
	L_1=\{ \omega^2, \omega^{10}, \omega^4, \omega^{671}, \omega^{474}, \omega^{616}, \omega^{454}, \omega^{401},
	\omega^{475}  \}.
$$

The quasi-cyclic representation of the elliptic quadrangle $Q(5,3)$ minus a spread, is given by a $4 \times 9$ matrix H$^{\mathrm{rep}}$. To compute the $(1,1)$ entry in H$^{\mathrm{rep}}$ (i.e. the top-left entry in the matrix), we determine those $i \in \{0,\ldots,27\}$ for which
$$
	a(\alpha^{i}x)^{4}-\alpha^{i}x+a^{9}=0,
$$
where $a= \omega^2$, $x= \omega^2$ and $\alpha=\omega^{26}$. It turns out that no such $i$ exists, so $\mathrm{H}^{\mathrm{rep}}_{x,a}$ is the empty set in this case. To determine the next entry in the matrix (i.e. the $(1,2)$ entry), we compute those $i \in \{0,\ldots,27\}$ for which
$$
	a(\alpha^{i}x)^{4}-\alpha^{i}x+a^{9}=0,
$$
where $a= \omega^{10}$, $x= \omega^2$ and $\alpha=\omega^{26}$. It turns out that both $i=9$ and $i=15$ work, so the subset is $\{9,15\}$. Continuing in this way, we determine that the matrix H$^{\mathrm{rep}}$ is
\begin{equation} \label{HrepQ(5,3)}
	\left(
	\begin{array}{ccccccccc}
			     & 9, 15 & 25     &        & 8, 12 & 1     & 16     &        & 1, 10 \\
			18   & 10    & 18, 28 & 3, 4   &       & 7, 19 & 14     &        &       \\
			1, 9 & 25    & 24     &        & 23    &       & 14, 16 & 14, 17 &       \\
			11   &       &        & 22, 27 & 15    & 6     &        & 15, 28 & 5, 16
		\end{array}
	\right).
\end{equation}
To construct the quasi-cyclic check matrix H from H$^{\mathrm{rep}}$, we substitute each empty set with a $b \times b$ matrix of zeros and each subset $\{ i\}$ by an identity matrix cyclically shifted to the right by $i$ bits. If there is more than one entry in the subset then we superimpose these cyclic shifts.

\begin{figure}[htbp]
	\centering
	\begin{center}
		\includegraphics[width=5in]{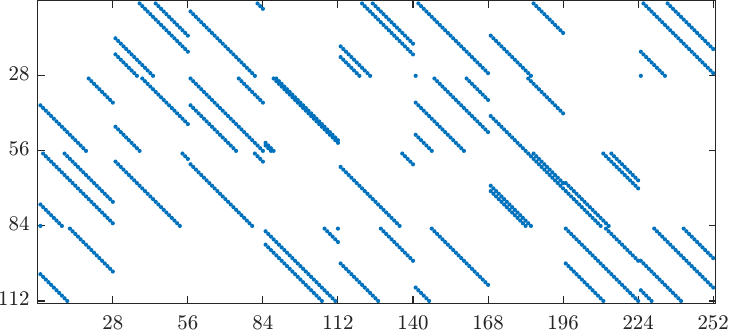}
	\end{center}
	\caption{The quasi-cyclic check matrix H for Q$(5,3)$. Dotted elements correspond to ones, and the rest are zeros.} \label{fig:H_Q53}
\end{figure}

Thus, the quasi-cyclic check matrix H for $q=3$ is the matrix in \cref{fig:H_Q53}.
We apply Gaussian elimination directly on H$^{\mathrm{rep}}$, to obtain a quasi-cyclic generator matrix in standard form for the code. As mentioned before,  it is not always the case that there exists a generator matrix in standard quasi-cyclic form, so we provide a generator matrix G for a large subcode $C'$ (called the implementable code) for which there is a generator matrix in standard quasi-cyclic form. Let $k$ be the dimension of the implementable code $C'$. As mentioned before equation (\ref{genmatrix}), the matrix G can be described by a $(k/b) \times ((n-k)/b)$ matrix P$^{\mathrm{rep}}$, where the $(i,j)$ entry of P$^{\mathrm{rep}}$ is $\mathrm{P}_{ij}$, a vector of $\{0,1\}^b$. We provide a matrix P$^{\mathrm{rep}}$, whose entries are the first row of a $b \times b$ circulant matrix. Recall that, replacing each first row with the full circulant $b \times b$ matrix, one obtains a matrix $\mathrm{P}$, where
$\mathrm{G} =(\mathrm{P}\ | \ \mathrm{id})$
is a generator matrix for the implementable code $C'$.

\begin{figure}
	\resizebox{\textwidth}{!}{%
		P$^{\mathrm{rep}} =
			\begin{pmatrix}
				0 0 0 0 0 0 0 1 1 0 1 1 0 1 0 1 1 1 0 0 0 0 0 0 1 0 0 0 & 1 0 0 1 0 0 1 0 0 1 1 1 0 1 0 1 0 0 1 0 1 1 1 0 0 0 1 0 & 1 1 1 0 1 0 1 0 0 1 0 0 0 0 1 0 0 1 0 0 1 1 1 1 0 0 1 1 & 1 0 0 0 0 0 0 0 0 0 0 0 0 0 0 0 0 0 0 0 0 0 0 1 1 0 1 1 \\
				0 0 0 0 0 1 0 0 0 1 1 0 0 0 1 1 0 0 0 0 0 0 0 0 0 0 0 0 & 1 0 1 0 1 0 0 1 0 0 0 0 0 1 1 1 1 1 1 1 0 0 0 0 0 1 0 1 & 1 0 1 1 1 1 1 0 0 1 1 0 1 1 1 1 1 1 1 0 1 0 0 0 0 1 1 1 & 0 0 0 0 0 0 0 0 0 0 0 0 0 0 0 0 0 0 0 0 0 0 0 0 0 1 1 0 \\
				0 0 0 0 0 0 1 0 1 1 0 0 1 1 1 0 1 1 0 0 0 0 0 0 0 0 0 0 & 0 1 1 0 0 0 0 1 1 0 0 1 0 1 0 0 0 0 0 0 0 0 0 0 1 0 0 0 & 0 1 0 0 0 0 1 0 0 1 0 1 0 0 0 0 0 0 0 0 0 1 0 0 1 0 0 1 & 1 0 0 0 0 0 0 0 0 0 0 0 0 0 0 0 0 0 0 0 0 0 1 0 1 1 0 1 \\
				0 0 0 0 0 0 0 1 1 1 0 1 1 1 1 0 0 0 0 0 0 0 0 0 1 0 0 0 & 1 0 0 1 1 0 0 1 0 1 1 1 0 1 1 0 0 1 1 0 1 0 1 1 1 0 1 0 & 1 1 0 0 1 0 0 0 0 0 0 1 1 1 0 0 1 1 0 0 0 1 0 0 1 1 1 0 & 0 0 0 0 0 0 0 0 0 0 0 0 0 0 0 0 0 0 0 0 0 0 0 1 1 1 0 0 \\
				0 0 0 0 0 0 1 1 1 1 1 0 0 1 1 1 0 1 0 0 0 0 0 1 0 0 0 0 & 1 0 1 1 0 0 1 1 1 1 1 1 0 0 1 0 0 0 1 1 0 1 0 0 0 0 0 1 & 1 1 1 1 0 0 1 1 0 0 1 0 1 1 0 0 0 1 0 0 1 1 1 1 1 1 1 1 & 1 0 0 0 0 0 0 0 0 0 0 0 0 0 0 0 0 0 0 0 0 0 0 1 1 1 1 0 \\
			\end{pmatrix}
		$
	}
	\begin{center}
		\includegraphics[width=5in]{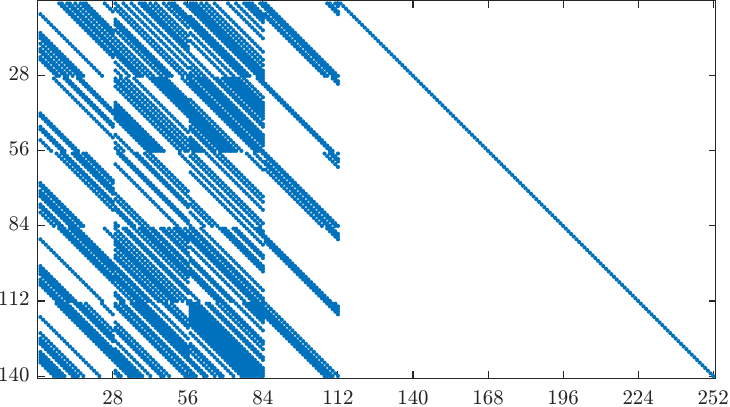}
	\end{center}
	\caption{The matrix P$^{\mathrm{rep}}$ (top) and corresponding quasi-cyclic generator matrix G for Q$(5,3)$ (bottom). Dotted elements correspond to ones, and the rest are zeros.}  \label{fig:G_Q_53}
\end{figure}

In the example for $q=3$ the matrix P$^{\mathrm{rep}}$ and the corresponding quasi-cyclic generator matrix G are in~\cref{fig:G_Q_53}.

%Thus, the quasi-cyclic generator matrix G is the matrix in \cref{fig:G_Q_53}.

The block size $b$ is $28$, so the check matrix H is a $112 \times 252$ matrix and the generator matrix G of the implementable code $C'$ is a $140 \times 252$ matrix. This $[252,140]$ code $C'$ is the code whose parameters appear in the first row in \cref{tableq5q}. We have constructed the code from the quasi-cyclic representation of the elliptic quadric quadrangle $\mathrm{Q}(5,3)$. Observe that it is fully described by the matrices H$^{\mathrm{rep}}$ and P$^{\mathrm{rep}}$.

\section{The quasi-cyclic representations of the classical finite generalized quadrangles.} \label{sec:gq}

As mentioned in the introduction, there are six classical finite generalized quadrangles, the elliptic quadric quadrangle $Q(5,q)$, the symplectic quadrangle $W(3,q)$, the Hermitian quadrangle $H(4,q^2)$ and their duals. In this section, we give a complete description of the quasi-cyclic representation of all finite classical generalized quadrangles. We explicitly detail how to calculate H$^{\mathrm{rep}}$ for each quadrangle from the quasi-cyclic representations of these quadrangles. As mentioned before, in the case of $Q(5,q)$ and $W(3,q)$, if we take all the lines except for a chosen spread then the block size can be increased significantly. We only need to describe the representation for the elliptic quadric quadrangle, the symplectic quadrangle and the Hermitian quadrangle $H(4,q^2)$ since the representation for their duals can be obtained by transposing the check matrix. In terms of H$^{\mathrm{rep}}$ this is accomplished in the following way. Let $\overline{\mathrm{H}^{\mathrm{rep}}}$ denote the H$^{\mathrm{rep}}$ matrix for the dual quadrangle. Suppose that the rows of  H$^{\mathrm{rep}}$ are indexed by the elements of $P_1$ and the columns of H$^{\mathrm{rep}}$ are indexed by the elements of $L_1$. Then the rows of $\overline{\mathrm{H}^{\mathrm{rep}}}$ are indexed
by the elements of $L_1$ and whose columns are indexed by the elements of $P_1$. For $x \in P_1$ and $a \in L_1$, the $(a,x)$ entry
of $\overline{\mathrm{H}^{\mathrm{rep}}}$ is the subset
$$
	\overline{\mathrm{H}^{\mathrm{rep}}_{a,x}}=\{-r \pmod{b} \ | \ r \in H_{x,a}^{\mathrm{rep}}\},
$$
where $b$ is the block size of the quasi-cyclic representation.

%\subsection{The elliptic quadric quadrangle $Q(5,q)$}
\subsection{The elliptic quadric quadrangle \texorpdfstring{$Q(5,q)$}{Q(5,q)}}

The elliptic quadric quadrangle $Q(5,q)$ is the point-line incidence structure whose points and lines are the $1$ and $2$-dimensional totally isotropic subspaces of an elliptic quadratic form defined over ${\mathbb F}_{q}^{6}$. The quasi-cyclic representation of $Q(5,q)$ has block size $(q^2-q+1)/3$ or $q^2-q+1$ depending on whether $3$ divides $q+1$. However, we can increase the block size to $b=q^3+1$ by taking a subset of the $q^5+q^3+q^2+1$ lines, obtained by removing a spread. This subset $\Sigma_A$, which is indexed by the elements of $L_A$, has size $q^5+q^2$.

Let
$$
	P=\{ x \in {\mathbb F}_{q^{6}} \ | \ 1+x^{q^3+1}+x^{(q^3+1)(q+1)}=0\},
$$
let
$$
	L_A=\{ a \in {\mathbb F}_{q^{6}} \ | \ 1+a^{q^3+1}+a^{q^4+q}+a^{q^5+q^2}=0\}
$$
and let
$$
	L_C=\{ c \in {\mathbb F}_{q^{6}} \ | \ 1+c^{q^2-q+1}+c^{q^3+1}=0\}.
$$

Let $\Sigma_C$ denote a set of $q^3+1$ lines of $Q(5,q)$, forming a spread, which will be explicitly given in the proof of \cref{thm:qrepelliptic}, and are indexed by the elements in $L_C$.

Let $P_1$ be a subset of $P$ of size $q+1$ such that for all $x_0,x_1 \in P_1$, $x_0^{q^3+1} \neq x_1^{q^3+1}$.

Let $L_1$ be a subset of $L_A$ of size $q^2$ with the property that for $a_0,a_1 \in L_1$, $a_0^{q^3+1} \neq a_1^{q^3+1}$.

The check matrix $\mathrm{H}$, for the incidence matrix of $Q(5,q) \setminus \Sigma_C$, is fully described by H$^{\mathrm{rep}}$, a $(q+1) \times q^2$ matrix whose rows are indexed by elements of $P_1$ and whose lines are indexed by elements of $L_1$. For $x \in P_1$ and $a \in L_1$, the $\mathrm{H}_{x,a}$ entry
of H$^{\mathrm{rep}}$ is a subset $\mathrm{H}^{\mathrm{rep}}_{x,a}$ of $\{0,\ldots,b-1 \}$, where $i \in \mathrm{H}^{\mathrm{rep}}_{x,a}$ indicates a $1$ in the $(i+1)$-st coordinate of the first row of the circulant $b \times b$ matrix indexed by $(x,a)$ in the quasi-cyclic matrix H.

Let $\alpha$ be a primitive $b$-th root of unity, where $b=q^3+1$.

\begin{theorem} \label{qcrepelliptic}
	The quasi-cyclic representation of $Q(5,q) \setminus \Sigma_C$ is given by $i \in \mathrm{H}^{\mathrm{rep}}_{x,a}$ for $x \in P_1$ and $a \in L_1$ if and only if
	$$
		a(\alpha^{i}x)^{q+1}-\alpha^{i}x+a^{q^2}=0,
	$$
	where $i \in \{0,\ldots,b-1\}$, and $b=q^3+1$.
\end{theorem}

\begin{proof}
	The elliptic quadric quadrangle $Q(5,q)$ is the point-line incidence structure whose points are the $1$-dimensional  totally isotropic subspaces of the quadric defined over ${\mathbb F}_{q^{6}}$ by
	\begin{equation} \label{qform}
		X^{q^3+1}+X^{q^4+q}+X^{q^5+q^2}
	\end{equation}
	and whose lines are the $2$-dimensional totally isotropic subspaces.

	One dimensional subspaces are given by
	$$
		X^{q}=xX
	$$
	for some $x \in {\mathbb F}_{q^{6}}$, which is necessarily a $(q-1)$-st power.

	Thus, the set of points of $Q(5,q)$ is the set
	$$
		P=\{ x \in {\mathbb F}_{q^{6}} \ | \ 1+x^{q^3+1}+x^{(q^3+1)(q+1)}=0\}.
	$$

	The lines $L$ are the totally isotropic two-dimensional subspaces. These split into two classes $\Sigma_A$ and $\Sigma_C$.

	The $(q^3+1)q^2$ lines in $\Sigma_A$, which are indexed by $a \in L_A$, are $\ell_{a}(X)=0$, where
	$$
		\ell_{a}(X)=aX^{q^2}-X^{q}-a^{q^2}X.
	$$

	To observe that $\ell_{a}(X)=0$ is totally isotropic, note that
	$$
		aX^{q^2}=X^{q}+a^{q^2}X
	$$
	implies
	$$
		a^{q+1}X^{q^3}=(1+a^{q^3+1})X^{q}+a^{q^2}X
	$$
	$$
		a^{q^2+q+1}X^{q^4}=-a^{q^5+q^2}X^{q}+a^{q^2}(1+a^{q^4+q})X
	$$
	$$
		a^{q^3+q^2+q+1}X^{q^5}=a^{q^5+q^3+q^2+1}X^{q}-a^{q^3+q^2+1}X
	$$
	and so, calculating modulo $\ell_a(X)$,
	$$
		a^{q^3+q^2+q+2}(X^{q^5+q^2}+X^{q^4+q}+X^{q^3+1})
	$$
	$$
		=(a^{q^5+q^3+q^2+1}X^{q}-a^{q^3+q^2+1}X)(X^{q}+a^{q^2}X)
	$$
	$$
		+a^{q^3+1}X^q(-a^{q^5+q^2}X^{q}+a^{q^2}(1+a^{q^4+q})X)
	$$
	$$
		+a^{q^3+q^2+1}X((1+a^{q^3+1})X^{q}+a^{q^2}X)
	$$
	$$
		=a^{q^3+q^2+1}X^{q+1}(a^{q^5+q^2}-1+(1+a^{q^4+q})+a^{q^3+1}+1)=0.
	$$

	Let $\alpha$ be a primitive $b$-th root of unity, where $b=q^3+1$.

	For $x \in P$ and $a \in L_A$,
	$$
		x \in \ell_a \Leftrightarrow ax^{q+1}-x+a^{q^2}=0
		\Leftrightarrow \alpha^{-q}a (\alpha x)^{q+1}-(\alpha x)+(\alpha^{-q}a)^{q^2}=0
		\Leftrightarrow \alpha x \in \ell_{\alpha^{-q} a}.
	$$

	Note that $(b,q)=1$, so we can order the points of the $(x,a)$ block as $\alpha^{i}x$ and lines as $\alpha^{-iq}a$, for $i$ from $0$ to $m-1$, and obtain a quasi-cyclic representation of block size $q^3+1$.

	The quasi-cyclic representation for $Q(5,q) \setminus \Sigma$ is then fully described by H$^{\mathrm{rep}}$, a $(q+1) \times q^2$ matrix whose rows are indexed by elements of $P_1$ and whose columns are indexed by elements of $L_1$. For $x \in P_1$ and $a \in L_1$, the $(x,a)$ entry H$^{\mathrm{rep}}_{x,a}$
	of H$^{\mathrm{rep}}$ is a subset of $\{0,\ldots,b-1\}$ which contains $i$ if and only if
	\begin{equation} \label{aeqnelliptic}
		a(\alpha^{i}x)^{q+1}-\alpha^{i}x+a^{q^2}=0.
	\end{equation}
\end{proof}

Let $m=q^2-q+1$ if $q \neq 2$ mod $3$ and $m=\frac{1}{3}(q^2-q+1)$ if $q=2$ mod $3$.

Let $P_2$ be a subset of $P$ of size $(q+1)(q^3+1)/m$ such that for all $y_0,y_1 \in P_2$, $y_0^{m} \neq y_1^{m}$.

Let $L_{2,C}$ be a subset of $L_C$ of size $(q^3+1)/m$ with the property that for $c_0,c_1 \in L_2$, $c_0^{m} \neq c_1^{m}$. Let $L_{2,A}$ be a subset of $L_A$ of size $(q^5+q^2)/m$ with the property that for $a_0,a_1 \in L_{2,A}$, $a_0^{m} \neq a_1^{m}$.

To obtain the full quasi-cyclic representation of $Q(5,q)$ we will construct a matrix $H^{\mathrm{rep}+}$, whose rows are indexed by the elements of $P_2$ and whose columns are indexed by the elements of $L_{2,C}$ and then the elements of $L_{2,A}$.

\begin{theorem} \label{thm:qrepelliptic}
	The quasi-cyclic representation of $Q(5,q)$ is given by $H^{\mathrm{rep}+}$, where $i \in \mathrm{H}^{\mathrm{rep}+}_{x,c}$ for $x \in P_2$ and $c \in L_{2,C}$ if and only if
	$$
		(\alpha^{i}x)^{q+1}-c=0,
	$$
	and
	$i \in \mathrm{H}^{\mathrm{rep}+}_{x,a}$ for $x \in P_2$ and $a \in L_{2,A}$ if and only if
	$$
		a(\alpha^{i}x)^{q+1}-\alpha^{i}x+a^{q^2}=0,
	$$
	where in both cases $i$ runs from $0$ to $m-1$, and $m=q^2-q+1$ if $q \neq 2 \mod 3$ and $m=\frac{q^2-q+1}{3}$ if $q=2 \mod 3$.
\end{theorem}

\begin{proof}
	That $i \in \mathrm{H}^{\mathrm{rep}+}_{x,a}$ if and only if $x \in P_2$ and $a \in L_{2,A}$ and
	$$
		a(\alpha^{i}x)^{q+1}-\alpha^{i}x+a^{q^2}=0,
	$$
	follows from (\ref{aeqnelliptic}).

	The $q^3+1$ lines in $\Sigma_C$ are $\ell_{c}(X)=0$, where
	$$
		\ell_{c}(X)=X^{q^2}-cX,
	$$
	where $c \in L_C$.

	We will prove that the lines $\ell_{c}(X):=X^{q^2}-cX$, where $c^{q^3+1}+c^{q^2-q+1}+1=0$, are totally isotropic.

	Observe that
	$$
		X^{q^2}=cX,
	$$
	implies
	$$
		X^{q^3}=c^qX^q, \ \ X^{q^4}=c^{q^2+1}X, \ \ X^{q^5}=c^{q^3+q}X^q
	$$
	and so
	$$
		X^{q^5+q^2}+X^{q^4+q}+X^{q^3+1}=(c^{q^3+q+1}+c^{q^2+1}+c^q)X^{q+1}=0.
	$$

	For  $x \in P$ and $c \in L_C$,
	$$
		x \in \ell_c \Leftrightarrow x^{q+1}-c=0  \Leftrightarrow \alpha x \in \ell_{\alpha^{-(q+1)} c}.
	$$

	The above equation implies that we can obtain a quasi-cyclic representation for $\mathrm{H}$ by letting $\alpha$ run through the $m$-th roots of unity, where $m$ must be chosen such that $(m,q+1)=1$. Recall that $m$ is a divisor of $q^3+1$. In the case that 3 does not divide $q-2$, we can choose $m=q^2-q+1$, and in the case that $3$ does divide $q-2$, we can set $m=\frac{1}{3}(q^2-q+1)$, as mentioned before. These block sizes coincide with those listed in \cite{LP2005}.

	For $x \in P_2$ and $c \in L_{2,C}$, the $(x,c)$ entry
	of H$^{\mathrm{rep}+}$ is $v(x,c)$ a subset of $\{1,\ldots,m\}$. As before, $i \in v(x,c)$ implies that to the $(i+1)$-th coordinate in the first row of the circulant $m \times m$ matrix indexed by $(x,c)$  has a $1$ in the quasi-cyclic matrix H$_2$. Thus, $i\in v(x,c)$ if and only if
	$$
		(\alpha^{i}x)^{q+1}-c=0,
	$$
	where $\alpha$ is a primitive $m$-th root of unity in ${\mathbb F}_{q^6}$, and $i \in \{0,\ldots,m-1\}$.
\end{proof}

\subsection{The symplectic quadrangle \texorpdfstring{$W(3,q)$}{W(3,q)}.}

The symplectic quadrangle $W(3,q)$ is obtained from the totally isotropic subspaces of a symplectic (or null or alternating) form defined on ${\mathbb F}_q^4$.
The quasi-cyclic representation of $W(3,q)$ has block size $q^2+1$ or $(q^2+1)/2$ depending on whether $q$ is even or odd. However, we can increase the block size to $b=q^2+1$ in all cases by removing a spread, denoted $\Sigma_C$ below.

Let $\gamma$ is a fixed element of ${\mathbb F}_{q^4}$ with the property that $\gamma^{q^2}=-\gamma$.

Let $\alpha$ be a fixed primitive $b$-th root of unity in ${\mathbb F}_{q^4}$.

Let
$$
	P=\{ x \in {\mathbb F}_{q^4} \ | \ x ^{(q^2+1)(q+1)}=1\},
$$
let
$$
	L_A=\{ a \in {\mathbb F}_{q^4} \ | \ \gamma^{1-q}a^{q(q^2+1)}-\gamma^{q-1} a^{q^2+1}+1=0\},
$$
and let
$$
	L_C=\{ c \in {\mathbb F}_{q^4} \ | \ c^{q^2+1}-1=0\},
$$
Let $\Sigma_C$ denote a set of $q^2+1$ lines of $W(3,q)$, forming a spread, which will be explicitly given in the proof of \cref{thm:qcrepsympletic}.

Let $P_1$ be a subset of $P$ of size $q+1$ such that for all $x_0,x_1 \in P_1$, $x_0^{q^2+1} \neq x_1^{q^2+1}$.

Let $L_1$ be a subset of $L_A$ of size $q$ with the property that for $a_0,a_1  \in L_1$, $a_0^{q^2+1} \neq a_1^{q^2+1}$.

The check matrix $\mathrm{H}$ for $W(3,q)\setminus \Sigma_C$ is fully described by H$^{\mathrm{rep}}$, a $(q+1) \times q$ matrix whose rows are indexed by elements of $P_1$ and whose columns are indexed by elements of $L_1$. For $x \in P_1$ and $a \in L_1$, the $(x,a)$ entry
of H$^{\mathrm{rep}}$ is a subset $\mathrm{H}^{\mathrm{rep}}_{x,a}$ of $\{0,\ldots,b-1\}$, where $i \in \mathrm{H}^{\mathrm{rep}}_{x,a}$ implies that there is a $1$ in the $(i+1)$-st coordinate of the first row of the circulant matrix indexed by $(x,a)$ in the quasi-cyclic matrix $\mathrm{H}$.

\begin{theorem} \label{thm:qcrepsympletic}
	The quasi-cyclic representation of $W(3,q) \setminus \Sigma_C$ is given by $i\in \mathrm{H}^{\mathrm{rep}}_{x,a}$ for $x \in P_1$ and $a\in L_1$ if and only if
	\begin{equation} \label{symplecticequation}
		a(\alpha^{i}x)^{q+1}+\alpha^{i}x-\gamma^{1-q}a^q=0,
	\end{equation}
	where $i\in \{0,\ldots,b-1\}$, and $b=q^2+1$.
\end{theorem}

\begin{proof}
	We will consider the vectors of ${\mathbb F}_{q}^4$ as elements of ${\mathbb F}_{q^4}$ and use the symplectic form
	$$
		\phi(X,Y):=\gamma X Y^{q^2}+\gamma^q X^q Y^{q^3}-\gamma X^{q^2}Y -\gamma^q X^{q^3} Y^{q},
	$$
	where $\gamma$ is a fixed element of ${\mathbb F}_{q^4}$ with the property that $\gamma^{q^2}=-\gamma$. Note that this choice of $\gamma$ ensures that $\phi(x,y) \in {\mathbb F}_q$ for all $x,y \in {\mathbb F}_{q^4}$.

	The points of $W(3,q)$ are the one-dimensional subspaces
	$$
		X^q=xX,
	$$
	where $x^{(q^2+1)(q+1)}=1$. Thus, we define $P$ to be the set
	$$
		P=\{ x \in {\mathbb F}_{q^4} \ | \ x ^{(q^2+1)(q+1)}=1\}.
	$$
	The $q^3+q$ lines of $\Sigma_A$ are the lines $\ell_a(X)=0$, where
	$$
		\ell_a(X)=aX^{q^2}+X^q-\gamma^{1-q} a^qX,
	$$
	and $a \in L_A$.

	To check that these are indeed lines it is enough to observe that
	$$
		a^{q+1}\ell_a(X)^{q^2}-a\ell_a(X)^q+\gamma^{q-1}a^{q^2+1}\ell_a(X)=a^{q^2+q+1}(X^{q^4}-X).
	$$
	To check that these lines are totally isotropic, it suffices to observe that if $X$ and $Y$ are joined by the line $\ell_a$, then
	$$
		(X^{q^2}Y-Y^{q^2}X)a=(Y^qX-X^qY)
	$$
	and so
	$$
		(X^{q^2}Y-Y^{q^2}X)^{q+1}\ell_a(X)=\gamma^{-q}(X^{q^2}Y^q-Y^{q^2}X^q)\phi(X,Y).
	$$
	Note that $X^{q^2}Y^q-Y^{q^2}X^q \neq 0$, since $a \neq 0$, and so $\phi(X,Y)=0$.

	Let $\alpha$ be a $b$-th -root of unity in ${\mathbb F}_{q^4}$, where $b=q^2+1$.

	For $x \in P$ and $a \in L_A$,
	$$
		x \in \ell_a \Leftrightarrow ax^{q+1}+x-\gamma^{1-q}a^q=0
		\Leftrightarrow (\alpha^{-q}a) (\alpha x)^{q+1}+\alpha x-\gamma^{1-q}(\alpha^{-q}a)^q=0
		\Leftrightarrow \alpha x \in \ell_{\alpha^{-q} a}.
	$$

	To obtain a quasi-cyclic representation of block size $q^2+1$ we choose $P_1$ to be a subset of $P$ of size $q+1$ such that for all $x_0,x_1 \in P_1$, $x_0^{q^2+1} \neq x_1^{q^2+1}$. Let $L_1$ be a subset of $L_A$ of size $q$ with the property that for $\ell_{a_0}$ and $\ell_{a_1} \in L_1$, $a_0^{q^2+1} \neq a_1^{q^2+1}$.

	The quasi-cyclic representation for $W(3,q) \setminus \Sigma_C$ is then fully described by H$^{\mathrm{rep}}$, a $(q+1) \times q$ matrix whose rows are indexed by elements of $P_1$ and whose columns are indexed by elements of $L_1$. For $x \in P_1$ and $a \in L_1$, the $(x,a)$ entry H$^{\mathrm{rep}}_{x,a}$
	of H$^{\mathrm{rep}}$ is a subset of $\{0,\ldots,b-1\}$ which contains $i$ if and only if
	$$
		a(\alpha^{i-1}x)^{q+1}+\alpha^{i-1}x-\gamma^{1-q}a^q=0.
	$$

\end{proof}

Let $m=q^2+1$ if $q$ is even and $m=\frac{1}{2}(q^2+1)$ if $q$ is odd.

Let $P_2$ be a subset of $P$ of size $(q+1)(q^2+1)/m$ such that for all $y_0,y_1 \in P_2$, $y_0^{m} \neq y_1^{m}$.

Let $L_{2,C}$ be a subset of $L_C$ of size $(q^2+1)/m$ with the property that for $c_0,c_1 \in L_2$, $c_0^{m} \neq c_1^{m}$. Let $L_{2,A}$ be a subset of $L_A$ of size $q(q^2+1)/m$ with the property that for $a_0,a_1 \in L_{2,A}$, $a_0^{m} \neq a_1^{m}$.

To obtain the full quasi-cyclic representation of $W(3,q)$ we will construct a matrix $H^{\mathrm{rep}+}$, whose rows are indexed by the elements of $P_2$ and whose columns are indexed by $L_2$, the elements of $L_{2,C}$ and then $L_{2,A}$.

\begin{theorem} \label{qrepsymplectic}
	The quasi-cyclic representation of $W(3,q)$ is given by $H^{\mathrm{rep}+}$, where $i \in \mathrm{H}^{\mathrm{rep}+}_{x,c}$ for $x \in P_2$ and $c \in L_{2,C}$ if and only if
	$$
		(\alpha^{i}x)^{q+1}-c=0,
	$$
	$i \in \mathrm{H}^{\mathrm{rep}+}_{x,a}$ for $x \in P_2$ and $a \in L_{2,A}$ if and only if
	\begin{equation} \label{aeqnsymplectic}
		a(\alpha^{i}x)^{q+1}+\alpha^{i}x-\gamma^{1-q}a^q=0,
	\end{equation}
	where in both cases $i$ runs from $0$ to $m-1$, and $m=q^2+1$ if $q$ is even and $m=(q^2+1)/2$ if $q$ is odd.
\end{theorem}

\begin{proof}
	That $i \in \mathrm{H}^{\mathrm{rep}+}_{x,a}$ if and only if $x \in P_2$ and $a \in L_{2,A}$ and
	$$
		a(\alpha^{i}x)^{q+1}+\alpha^{i}x-\gamma^{1-q}a^q=0,
	$$
	follows from (\ref{aeqnsymplectic}).

	The $q^2+1$ lines of $\Sigma_C$ are $\ell_c(X)=0$, where
	$$
		\ell_c(X)=X^{q^2}-cX,
	$$
	and $c^{q^2+1}=1$, i.e. $c \in L_C$.

	To check that these are indeed lines it is enough to observe that

	$$
		\ell_c(X)^{q^2}+c^{q^2}\ell_c(X)=X^{q^4}-X.
	$$

	If $X$ and $Y$ are joined by the line $\ell_c$ then
	$$
		X^{q^2}Y-Y^{q^2}X=0
	$$
	and so $\phi(X,Y)=0$.

	For  $x \in P$ and $\ell_c \in L_C$,
	$$
		x \in \ell_c \Leftrightarrow x^{q+1}-c=0  \Leftrightarrow \alpha x \in \ell_{\alpha^{-(q+1)} c}.
	$$

	The above equation implies that we can complete the quasi-cyclic representation for $W(3,q)$ by letting $\alpha$ run through the $m$-th roots of unity, where $m$ must be chosen such that $(m,q+1)=1$. In the case $q$ is even, we can choose $m=q+1$, and in the case that $q$ is odd, we can set $m=\frac{1}{2}(q+1)$.

	For $x \in P_2$ and $c \in L_{2,C}$, the $(x,c)$ entry
	of H$^{\mathrm{rep}+}$ is $v(x,c)$ a subset of $\{1,\ldots,m\}$. As before, $i \in v(x,c)$ implies that the $(i+1)$-th coordinate in the first row of the circulant $m \times m$ matrix indexed by $(x,c)$  has a $1$ in the quasi-cyclic check matrix. Thus, $i\in v(x,c)$ if and only if
	$$
		(\alpha^{i}x)^{q+1}-c=0,
	$$
	where $\alpha$ is a primitive $m$-th root of unity in ${\mathbb F}_{q^4}$, and $i \in \{0,\ldots,m-1\}$.
\end{proof}

\subsection{The Hermitian quadrangle \texorpdfstring{$H(4,q^2)$}{H(4,q2)}}

The quadrangle $H(4,q^2)$ is the point-line incidence structure whose points are the $1$ and $2$-dimensional (over ${\mathbb F}_{q^2}$) totally isotropic subspaces of  the Hermitian surface defined over ${\mathbb F}_{q^{2}}^5$. There is a quasi-cyclic representation of $H(4,q^2)$ with block size $b=(q^5+1)/(q+1)$, which we will describe explicitly.

Let $\alpha$ be a fixed primitive $b$-th root of unity in ${\mathbb F}_{q^{10}}$.

Let
$$
	P=\{ x \in {\mathbb F}_{q^{10}} \ | \ 1+x^{q^5+1}+x^{(q^5+1)(q^2+1)}+x^{(q^5+1)/(q+1)}+x^{(q^2+q+1)(q^5+1)/(q+1)}=0\}.
$$
Define
$$
	\Delta(c)=(1+c)^{q^2+1}+c^q,
$$
and
$$
	\Omega(c)=(1+c)^{q^4+1}+c^{q^4-q^2+1}.
$$

Let
$$
	L'=\{ (c,a) \ | \ \Delta(c)\neq 0, \ \Omega(c)=0, \ a^{(q^5+1)/(q+1)}=\frac{c(c^{q^2+q^4}+c^{q^2}-c^{q^4+q^3-q}-c^{q^4+q^3})}{(c^{q^4+q}+c^q+c^{q^4+q^3+q}+c^{q+q^3}+c^{q^3+q^4})}\}
$$
\begin{lemma} \label{sizeL'}
	The set $L'$ has size at most $(q^3-q^2)(q^5+1)$.
\end{lemma}

\begin{proof}
	Let $f(c)=\Delta(c-1)$. One can check that
	$$
		c^{q^2+q}f^{q^3}-c^qf^{q^2}+(1-c^{q^2+1})f^q+(c^{q^2}-1)f=0 \pmod{c^{q^5}-c},
	$$
	which implies that $\Delta$ has $q^2+1$ roots, all of which are in ${\mathbb F}_{q^5}$.

	Furthermore, if $\Delta(c)=0$ then
	$$
		(1+c)^{q^2+1}=-c^q
	$$
	which implies
	$$
		(1+c)^{q^6+1}=-c^{q(q^4-q^2+1)}
	$$
	and (raising this to the $q^4$-th power),
	$$
		(1+c)^{q^4+1}=-c^{q^4-q^2+1},
	$$
	so $c$ is a root of $\Omega$.

	Therefore, there are at most $q^4-q^2$ solutions to $\Omega(c)=0$, $\Delta(c) \neq 0$ and each one of these provides at most $(q^5+1)/(q+1)$ possible solutions for $a$.

\end{proof}

Let
$$
	L''=\{ (c,a) \ | \ \Delta(c)=0, \ a^{q^5+1}+c^{-q^2+q}a^{(q^5+1)/(q+1)}-c=0 \}.
$$

\begin{lemma} \label{sizeL''}
	The set $L''$ has size at most $(q^2+1)(q^5+1)$.
\end{lemma}

\begin{proof}
	This is immediate since $\Delta(c)$ has degree $q^2+1$ and for a fixed $c$
	$$
		a^{q^5+1}+c^{-q^2+q}a^{(q^5+1)/(q+1)}-c=0
	$$
	has at most $q^5+1$ solutions for $a$.
\end{proof}

We will see later that equality holds in both \cref{sizeL',sizeL''}.

Let $L=L' \cup L''$ and note that equality in both \cref{sizeL',sizeL''} implies $|L|=(q^3+1)(q^5+1)$.

Let $P_1$ be a subset of $P$ of size $(q+1)(q^2+1)$ such that for all $x_0,x_1 \in P_1$, $x_0^{(q^5+1)/(q+1)} \neq x_1^{(q^5+1)/(q+1)}$.

Let $L_1$ be a subset of $L$ of size $(q+1)(q^3+1)$ with the property that for $(a_0,c)$ and $(a_1,c) \in L_1$, $a_0^{(q^5+1)/(q+1)} \neq a_1^{(q^5+1)/(q+1)}$.

The incidence matrix of the quadrangle $\mathrm{H}$ is then fully described by H$^{\mathrm{rep}}$, a $$
	(q+1)(q^2+1)\times  (q+1)(q^3+1)
$$
matrix whose columns are indexed by elements of $L_1$ and whose rows are indexed by elements of $P_1$. For $x \in P_1$ and $(a,c) \in L_1$, the $(x,(a,c))$ entry
of H$^{\mathrm{rep}}$ is a subset $\mathrm{H}^{\mathrm{rep}}_{x,(a,c)}$ of $\{0,\ldots,b-1\}$, where $i \in \mathrm{H}^{\mathrm{rep}}_{x,(a,c)}$ implies that the $(i+1)$-coordinate of the first row of the circulant matrix indexed by $(x,(a,c))$ in the quasi-cyclic matrix $\mathrm{H}$ is $1$.

\begin{theorem}
	The quasi-cyclic representation of $H(4,q^2)$ is given by $i \in \mathrm{H}^{\mathrm{rep}}_{x,(a,c)}$ for $x \in P_1$ and $(a,c) \in L_1$ if and only if
	\begin{equation} \label{hermitianequation}
		(\alpha^{i}x)^{q^2+1}-a^{q^2}\alpha^{i}x-a^{q^2+1}c^{-1}=0,
	\end{equation}
	where $i   \in \{0,\ldots,b-1\}$, and $b=(q^5+1)/(q+1)$.
\end{theorem}

\begin{proof}

	Define $H(4,q^2)$ as the point-line incidence structure whose points are the $1$-dimensional (over ${\mathbb F}_{q^2}$) totally isotropic subspaces of  the Hermitian surface defined by
	\begin{equation} \label{hform}
		X^{q^5+1}+X^{q^7+q^2}+X^{q^9+q^4}+X^{q+q^6}+X^{q^3+q^8}
	\end{equation}
	and whose lines are the $2$-dimensional totally isotropic subspaces.

	One dimensional subspaces are given by
	$$
		X^{q^2}=xX
	$$
	for some $x \in {\mathbb F}_{q^{10}}$, which is necessarily a $(q^2-1)$-st power.

	Thus, we have a totally isotropic subspace if, substituting in (\ref{hform})
	$$
		1+x^{q^5+1}+x^{q^7+q^5+q^2+1}+x^{q^4-q^3+q^2-q+1}+x^{q^6+q^4-q^3+q^2+1}=0.
	$$
	Substituting $y=x^{(q^5+1)/(q+1)}$, this becomes
	$$
		1+y+y^{q+1}+y^{q^2+q+1}+y^{q^3+q^2+q+1}=0.
	$$

	Two dimensional subspaces are given by
	\begin{equation} \label{tilines}
		X^{q^4}=aX^{q^2}+bX,
	\end{equation}
	for some $a,b \in {\mathbb F}_{q^{10}}$.

	Reducing modulo (\ref{tilines}),
	$$
		X^{q^6}=(a^{q^2+1}+b^{q^2})X^{q^2}+a^{q^2}bX,
	$$
	and
	$$
		X^{q^8}=(a^{q^4+q^2+1}+ab^{q^4}+a^{q^4}b^{q^2})X^{q^2}+(a^{q^4+q^2}b+b^{q^4+1})X.
	$$
	Substituting in (\ref{hform}) implies (\ref{tilines}) is totally isotropic if and only if
	$$
		X^{q^3+q^2} (a^{q^3+q}+b^{q^3}+a^{q^5+q^3+q+1}+a^{q+1}b^{q^5}+a^{q^5+1}b^{q^3}+a^{q^4+q^2+1}+ab^{q^4}+a^{q^4}b^{q^2})+
	$$
	$$
		X^{q^3+1} (a^{q}+a^{q^5}b^{q^3+1}+a^{q^5+q^3+q}b+a^qb^{q^5+1}+a^{q^4+q^2}b+b^{q^4+1})+
	$$
	$$
		X^{q^2+q} (a^{q^3}b^{q}+a^{q^5+q^3+1}b^q+a b^{q^5+q}+a^{q^2+1}+b^{q^2})+
	$$
	$$
		X^{q+1} (b^{q}+b^{q^5+q+1}+a^{q^2}b+a^{q^5+q^3}b^{q+1})=0.
	$$

	Substituting $b^{q^2}=a^{q^2+1}c^{-1}$ and $d=a^{(q^5+1)/(q+1)}$ implies that the coefficients of $X^{q+1}$, $X^{q^2+q}$, $X^{q^3+1}$ and $bX^{q^3+q^2}-aX^{q^3+1}$ are respectively
	$$
		d^{q^3+q^2+q+1}(1+c^{-q^{5}})+dc^q+c=0,
	$$
	$$
		d^{q^2+q+1}(1+c^{-q^{4}})+d+c^{1-q}+c=0,
	$$
	$$
		d^{q^3+q^2+q+1}(1+c^{-q^{5}}+c^{-q^{3}})+d^{q^2+q+1}(1+c^{-q^4})+c=0,
	$$
	$$
		d^{q^2+q+1}c^{-q^2}+(1+c^{-q^{3}})d^{q+1}-c=0.
	$$
	These equations reduce to
	\begin{equation} \label{cdeqn}
		d^{q+1}+c^{-q^2+q}d-c=0,
	\end{equation}
	\begin{equation} \label{lineard}
		(c^{q^3+1}+c+c^{q^3+q^2+1}+c^{q^2+1}+c^{q^3+q^2})^qd=(c^{q^3+q}+c^q-c^{q^3+q^2}-c^{q^3+q^2-1})^qc
	\end{equation}
	\begin{equation} \label{ceqn}
		(1+c)^{q^4+1}+c^{q^4-q^2+1}=0.
	\end{equation}
	One can check that (\ref{cdeqn}) and (\ref{lineard}) imply (\ref{ceqn}).

	A totally isotropic line is, therefore, using (\ref{tilines}),
	$$
		x^{q^2+1}-ax-a^{1+q^{-2}}c^{-q^{-2}}=0,
	$$
	where, as before, $x=X^{q^2-1}$, and $a^{(q^5+1)/(q+1)}=d$ where $c$ and $d$ must satisfy (\ref{cdeqn}), (\ref{lineard})  and (\ref{ceqn}).

	Let
	$$
		\theta(c)=c+c^{q^2+1}+c^{q^3+1}+c^{q^3+q^2}+c^{q^3+q^2+1}.
	$$
	One can check that
	\begin{equation} \label{deltathetaeqn}
		\Delta^q(c)-(\theta(c)/c)=c^{q^3+q}+c^q-c^{q^3+q^2}-c^{q^3+q^2-1}
	\end{equation}
	and that
	$$
		(c^q+1)\theta(c)=(c+c^{q^2}+c^{q^2+1})\Delta(c)^q-c^{q^2}\Delta(c).
	$$
	By (\ref{ceqn}), $\Omega(c)=0$.

	If $\theta(c)=0$ then (\ref{lineard}) and (\ref{deltathetaeqn}) imply that $\Delta(c)=0$ and by (\ref{cdeqn}), $(a,c) \in L''$.

	If $\theta(c) \neq 0$ then $\Delta(c) \neq 0$ and $(a,c) \in L'$.

	We have proven that all totally isotropic lines are indexed by $(a,c)$ which belongs to either in $L'$ or $L''$. Since, by~\cref{sizeL',sizeL''},
	$$
		|L'|+|L''|\leqslant (q^3+1)(q^5+1),
	$$
	and there are $ (q^3+1)(q^5+1)$ totally isotropic lines, we must have equality and that these are precisely all the indexes of the totally isotropic lines.

	Recall that $\alpha$ is a primitive $(q^5+1)/(q+1)$-st root of unity.

	The quasi-cyclic representation for $H(4,q^2)$ is then fully described by H$^{\mathrm{rep}}$,  whose rows are indexed by elements of $P_1$ and whose columns are indexed by elements of $L_1$. For $x \in P_1$ and $(a,c) \in L_1$, the $(x,(a,c))$ entry H$^{\mathrm{rep}}_{x,a}$
	of H$^{\mathrm{rep}}$ is a subset of $\{0,\ldots,b-1\}$ which contains $i$ if and only if
	$$
		(\alpha^{i}x)^{q^2+1}-a^{q^2}\alpha^{i}x-a^{q^2+1}c^{-1}=0.
	$$
	Note that replacing $a$ by $a^{q^2}$ and $c$ by $c^{q^2}$ does not alter the conditions on $a$ and $c$. We make this substitution for purely aesthetic reasons.

\end{proof}

\begin{table}[htbp]
	%\footnotesize
	\centering
	\caption{The block size, length, and complexity of constructing H$^{\mathrm{rep}}$ and P$^{\mathrm{rep}}$.} \label{table1}
	\begin{tabular}{|c|c|c|c|c|c|c|c|}
		\hline
		\multirow{2}{*}{Code} & increased           & \multirow{2}{*}{block size}            & approximate & complexity         & complexity         & minimum          & approximate               \\
		                      & block size $b$      &                                        & size $n$    & H$^{\mathrm{rep}}$ & P$^{\mathrm{rep}}$ & distance         & rate                      \\ \hline
		                      &                     &                                        &             &                    &                    &                  &                           \\
		$W(3,q)$              & $q^2+1$             & $q^2+1$ ($q$ even)                     & $q^3$       & $O(q^4)$           & $O(q^9)$           & $\geqslant 2q$   & $1-q^{-0.286}$ ($q$ even) \\
		                      &                     & $\frac{1}{2}(q^2+1)$ ($q$ odd)         &             &                    &                    &                  & 0.5 ($q$ odd)             \\

		$W(3,q)$              & $q^2+1$             & $q^2+1$ ($q$ even)                     & $q^3$       & $O(q^4)$           & $O(q^9)$           & $\geqslant 2q$   & $1-q^{-0.286}$ ($q$ even) \\
		dual                  &                     & $\frac{1}{2}(q^2+1)$ ($q$ odd)         &             &                    &                    &                  & 0.5 ($q$ odd)             \\
		                      &                     &                                        &             &                    &                    &                  &                           \\

		$Q(5,q)$              & $q^3+1$             & $q^2-q+1$ ($q=0,1$ mod $3$)            & $q^5$       & $O(q^6)$           & $O(q^{13})$        & $\geqslant 2q$   & $1-q^{-1}$                \\
		                      &                     & $\frac{1}{3}(q^2-q+1)$ ($q=2$ mod $3$) &             &                    &                    &                  &                           \\

		$Q(5,q)$              & $q^3+1$             & $q^2-q+1$ ($q=0,1$ mod $3$)            & $q^4$       & $O(q^6)$           & $O(q^{14})$        & $\geqslant q^3$  & $q^{-1}$                  \\
		dual                  &                     & $\frac{1}{3}(q^2-q+1)$ ($q=2$ mod $3$) &             &                    &                    &                  &                           \\
		                      &                     &                                        &             &                    &                    &                  &                           \\
		$H(4,q^2)$            & $\frac{q^5+1}{q+1}$ & $\frac{q^5+1}{q+1}$                    & $q^8$       & $O(q^{11})$        & $O(q^{22})$        & $\geqslant 2q^2$ & $1-q^{-1}$                \\

		                      &                     &                                        &             &                    &                    &                  &                           \\
		$H(4,q^2)$            & $\frac{q^5+1}{q+1}$ & $\frac{q^5+1}{q+1}$                    & $q^7$       & $O(q^{11})$        & $O(q^{23})$        & $\geqslant q^5$  & $q^{-1}$                  \\
		dual                  &                     &                                        &             &                    &                    &                  &                           \\
		\hline
	\end{tabular}
\end{table}

\section{The quasi-cyclic representation of the projective and affine spaces} \label{sectionpanda}

The same technique of embedding the geometry in a finite field extension can be used to construct the quasi-cyclic (which are sometimes cyclic) representations of the classical projective and affine spaces.

\subsection{The projective space \texorpdfstring{$\mathrm{PG}(k-1,q)$}{PG(k-1,q)}.}

The  projective space $\mathrm{PG}(k-1,q)$ is the geometry whose $j$-dimensional subspaces are the $(j-1)$-dimensional subspaces of ${\mathbb F}_q^k$. We will consider this vector space as the finite field ${\mathbb F}_{q^k}$.

Let $\alpha$ be a primitive $b=(q^k-1)/(q-1)$-th root of unity in ${\mathbb F}_{q^k}$.

\begin{theorem} \label{pgthm}
	The cyclic representation of the point-hyperplane incidences of $\mathrm{PG}(k-1,q)$ is given by a $1 \times 1$ matrix H$^{\mathrm{rep}}$, where the $(1,1)$ entry H$^{\mathrm{rep}}_{1,1}$
	of H$^{\mathrm{rep}}$ is a subset of $\{0,\ldots,b-1\}$, with $b=(q^k-1)/(q-1)$,  which contains $i$ if and only if
	$$
		\sum_{j=0}^{k-1} \alpha^{i(q^j-1)/(q-1)}=0.
	$$
\end{theorem}

\begin{proof}
	The points of $\mathrm{PG}(k-1,q)$ are the one-dimensional subspaces of ${\mathbb F}_{q^k}$, which are
	$$
		X^{q}=xX
	$$
	for some $x \in {\mathbb F}_{q^{k}}$, which is necessarily a $(q-1)$-st power.

	The hyperplanes of
	$PG(k-1,q)$ are the hyperplanes of ${\mathbb F}_{q^k}$, which are
	$$
		\sum_{j=0}^{k-1} (AX)^{q^j}=0,
	$$
	where $A \neq 0$.

	Let
	$$
		P=\{ x \in {\mathbb F}_{q^k} \ | \ x ^{(q^k-1)(q-1)}=1\},
	$$
	and let
	$$
		L=\{ a \in {\mathbb F}_{q^k} \ | \ a ^{(q^k-1)(q-1)}=1\}.
	$$
	Then, according to the above, the point-hyperplane incidence is given by the point $x\in P$ is incident with the hyperplane $\pi_a$, where $a \in L$, if and only if
	$$
		\sum_{j=0}^{k-1} (ax)^{(q^j-1)/(q-1)}=0.
	$$

	The cyclic representation of the point-hyperplane incidence has block size $b=(q^k-1)/(q-1)$ and is given by the following equation,
	\begin{equation} \label{projeqn}
		x \in \pi_{\alpha a} \Leftrightarrow  \sum_{j=0}^{k-1} (\alpha ax)^{(q^j-1)/(q-1)}=0 \Leftrightarrow
		\alpha x \in \pi_{a}.
	\end{equation}

	The cyclic representation for the point-hyperplane incidences of $\mathrm{PG}(k-1,q)$ is then fully described by a $1 \times 1$ matrix H$^{\mathrm{rep}}$, where the $(1,1)$ entry H$^{\mathrm{rep}}_{1,1}$
	of H$^{\mathrm{rep}}$ is a subset of $\{0,\ldots,b-1\}$ which contains $i$ if and only if
	$$
		\sum_{j=0}^{k-1} (\alpha^i)^{(q^j-1)/(q-1)}=0.
	$$
\end{proof}

The line $\ell_{a,c}$ of $\mathrm{PG}(k-1,q)$ is the two-dimensional subspace of ${\mathbb F}_{q^k}$, defined by
$$
	X^{q^2}+aX^q+cX=0
$$
for some $a,c \in {\mathbb F}_{q^k}$ such that $X^{q^2}+aX^q+cX$ divides $X^{q^k}-X$.

This last condition is equivalent to $\mathrm{rank}(\mathrm{M}_{a,c}) \leqslant k-2$ where $\mathrm{M}_{a,c}$ is the $(k-1) \times k$ matrix
$$
	\mathrm{M}_{a,c}=\left(\begin{array}{ccccccc}
			0           & \ldots      & 0           & 0      & 1      & a   & c      \\
			0           & \ldots      & 0           & 1      & a^q    & c^q & 0      \\
			\vdots      &             & \adots      & \adots &        &     & \vdots \\
			\vdots      & \adots      & \adots      & \adots &        &     & \vdots \\
			\vdots      & \adots      & \adots      &        &        &     & \vdots \\
			1           & a^{q^{k-3}} & c^{q^{k-3}} & 0      & \ldots & 0   & 0      \\
			a^{q^{k-2}} & c^{q^{k-2}} & 0           & 0      & \ldots & 0   & 1      \\
		\end{array}
	\right)
$$

We have that
\begin{equation} \label{projlineeqn}
	\alpha x \in \ell_{ a,c} \Leftrightarrow  (\alpha x)^{q+1}+a \alpha x +c =0 \Leftrightarrow
	x \in \ell_{\alpha^{-q}a,\alpha^{-q-1}c},
\end{equation}
where $\alpha$ is again a primitive $b=(q^k-1)/(q-1)$-th root of unity in ${\mathbb F}_{q^k}$.

Since the non-zero roots of
$$
	X^{q^2}+aX^q+cX=0
$$
are $(q-1)$-st powers we have that $c^b=1$.

If $k$ is odd then $\mathrm{gcd}(b,q+1)=1$ so for any fixed $c$ we obtain an orbit of $b$ lines. Thus, we can set
$$
	L_1=\{ (a,1) \ | \ \mathrm{rank} (\mathrm{M}_{a,1} )\leqslant k-2 \}.
$$

The point-line incidence matrix for $\mathrm{PG}(k-1,q)$ is fully described by $\mathrm{H}^{\mathrm{rep}}$, a $1 \times (q^{k-1}-1)/(q^2-1)$ matrix whose columns are indexed by $L_1$.

We have the following theorem.

\begin{theorem} \label{replinesprojodd}
	The quasi-cyclic representation of the point-line incidences of $\mathrm{PG}(k-1,q)$, $k$ odd, is given by $i \in \mathrm{H}^{\mathrm{rep}}_{1,(a,1)}$ for $a \in L_1$ if and only if
	$$
		\alpha^{i(q+1)}+a \alpha^i  +1 =0
	$$
	where $i \in \{0,\ldots,b-1\}$ and $b=(q^k-1)/(q-1)$.
\end{theorem}

\begin{proof}
	We can put $x=c=1$ in \eqref{projlineeqn} and then $\alpha^i \in \ell_{a,1}$ if and only if
	$$
		\alpha^{i(q+1)}+a \alpha^i  +1 =0.
	$$
\end{proof}

If $k$ is even then we do not get a full orbit of lines if $a=0$ and $c^{(q^k-1)/(q^2-1)}=1$. Note that $\ell_{a_1,c_1}$ and $\ell_{a_2,c_2}$ are in the same orbit if there is an $\alpha$ such that $\alpha^{(q^{k}-1)/(q-1)}=1$ and $a_2=\alpha^{-q}a_1$ and $c_2=\alpha^{-q-1} c_1$.

Thus, we must choose a set of representatives $P_2$ and $L_2$ for the point and line orbits respectively, where $b=(q^{k}-1)/(q^2-1)$.

The point-line incidence matrix for $\mathrm{PG}(k-1,q)$ is fully described by $\mathrm{H}^{\mathrm{rep}}$, a $(q+1) \times (q^{k-1}-1)/(q-1)$ matrix whose rows are indexed by $P_2$ and whose columns are indexed by $L_2$, where
$$
	P_2 =\{ x \ | \ x^{q+1}=1 \}
$$
and a set $L_2$ of size $(q^{k-1}-1)/(q-1)$ such that $(0,1) \in L_2$ and for distinct $(a_1,c_1), (a_2,c_2) \in L_2$,
$$
	\mathrm{rank} (\mathrm{M}_{a,c}) \leqslant k-2
$$
and not both $a_1^{(q^k-1)/(q-1)}=a_2^{(q^k-1)/(q-1)}$ and $c_1^{(q^k-1)/(q^2-1)}=c_2^{(q^k-1)/(q^2-1)}$.

\begin{theorem} \label{replinesprojeven}
	The quasi-cyclic representation of the point-line incidences of $\mathrm{PG}(k-1,q)$, $k$ even, is given by $i \in \mathrm{H}^{\mathrm{rep}}_{x,(a,c)}$ for $(a,c) \in L_2$ if and only if
	$$
		(\alpha^{i}x)^{q+1}+a \alpha^i x +c =0,
	$$
	where $i \in \{0,\ldots,b-1\}$ and $b=(q^k-1)/(q^2-1)$.
\end{theorem}

\begin{proof}
	We can put $x\in P_2$ and $(a,c) \in L_2$ in \eqref{projlineeqn} and then $\alpha^i x \in \ell_{a,c}$ if and only if
	$$
		(\alpha^{i}x)^{q+1}+a \alpha^i x +c =0.
	$$
\end{proof}

As in the case of the elliptic and symplectic quadrangle, we can remove a spread of lines and increase the block size. Let $\Sigma$ be the spread of $(q^k-1)/(q^2-1)$ lines defined by
$$
	x^{q+1}-c=0,
$$
where $c^{(q^k-1)/(q^2-1)}=1$.

For any other fixed $c$ we obtain an orbit of $b$ lines. Thus, we find a set $L_3$ of size $(q^{k-1}-1)/(q^2-1)$ such that for distinct $(a_1,c_1), (a_2,c_2) \in L_3$,
$$
	\mathrm{rank} \mathrm{M}_{a,c} \leqslant k-2
$$
and not both $a_1^{(q^k-1)/(q-1)}=a_2^{(q^k-1)/(q-1)}$ and $c_1^{(q^k-1)/(q-1)}=c_2^{(q^k-1)/(q-1)}$.

The point-line incidence matrix for $\mathrm{PG}(k-1,q)\setminus \Sigma$ is fully described by $\mathrm{H}^{\mathrm{rep}}$, a $1 \times (q^{k-1}-1)/(q^2-1)$ matrix whose columns are indexed by $L_3$.

\begin{theorem} \label{replinesprojeven2}
	The quasi-cyclic representation of the point-line incidences of $\mathrm{PG}(k-1,q)\setminus \Sigma$, $k$ even, is given by $i \in \mathrm{H}^{\mathrm{rep}}_{a,c}$ for $(a,c) \in L_3$ if and only if
	$$
		\alpha^{i(q+1)}+a \alpha^i  +c =0,
	$$
	where $i \in \{0,\ldots,b-1\}$ and $b=(q^k-1)/(q-1)$.
\end{theorem}

\begin{proof}
	We can put $x=1$ and $(a,c) \in L_3$ in \eqref{projlineeqn} and then $\alpha^i \in \ell_{a,c}$ if and only if
	$$
		\alpha^{i(q+1)}+a \alpha^i  +c =0.
	$$
\end{proof}

\subsection{The affine space \texorpdfstring{$\mathrm{AG}(k,q)$}{AG(k,q)}.}

The affine space $\mathrm{AG}(k,q)$ is the geometry whose $j$-dimensional subspaces are the ${\mathbb F}_q$-cosets of the $j$-dimensional subspaces of ${\mathbb F}_q^k$. We will again consider this vector space as the finite field ${\mathbb F}_{q^k}$.

Let $\alpha$ be a fixed primitive $b=(q^k-1)/(q-1)$-th root of unity in ${\mathbb F}_{q^k}$.

We will describe the quasi-cyclic representation for the point-hyperplane incidences with block sizes of size $b$ by a $1 \times q$ matrix H$^{\mathrm{rep}}$ whose columns are indexed by the elements of ${\mathbb F}_q$.

\begin{theorem} \label{agthm}
	The quasi-cyclic representation of the point-hyperplane incidences of $\mathrm{AG}(k,q)$ is given by a $1 \times q$ matrix $\mathrm{H}^{\mathrm{rep}}$, where the $(1,a)$ entry $\mathrm{H}^{\mathrm{rep}}_{1,a}$
	of $\mathrm{H}^{\mathrm{rep}}$ is a subset of $\{0,\ldots,b-1\}$ which contains $i$ if and only if
	$$
		\sum_{j=0}^{k-1} \alpha^{iq^j}=a,
	$$
	where $b=(q^k-1)/(q-1)$ and $a \in  {\mathbb F}_q$.
\end{theorem}

\begin{proof}
	The points of $\mathrm{AG}(k,q)$ are the elements of ${\mathbb F}_q^k$ which we will consider as elements of $x \in {\mathbb F}_{q^k}$. Then the hyperplanes $\pi_{a,c}$ are defined by
	$$
		\sum_{j=0}^{k-1} (cx)^{q^j}=a,
	$$
	for some $a,c$. Since for all non-zero $\lambda \in {\mathbb F}_q$, the hyperplane defined by $(a,c)$ is the same as the hyperplane $(\lambda a, \lambda c)$ we can suppose $c^b=1$.

	For $\alpha$, a primitive $b$-th root of unity,
	\begin{equation} \label{affineeqn}
		x \in \pi_{a,\alpha c} \Leftrightarrow  \sum_{j=0}^{k-1} (\alpha cx)^{q^j}=a \Leftrightarrow
		\alpha x \in \pi_{a,c}.
	\end{equation}
	Thus, we can obtain an orbit of size $b$ by setting $c=x=1$ and taking the hyperplanes $\pi_{a,1}$ as representatives for each orbit, where $a \in {\mathbb F}_q$.

	Thus, we index the columns of H$^{\mathrm{rep}}$  by the elements of $a\in {\mathbb F}_q$ where the
	H$^{\mathrm{rep}}_{1,a}$
	of H$^{\mathrm{rep}}$ is a subset of $\{0,\ldots,b-1\}$ which contains $i$ if and only if
	$$
		\sum_{j=0}^{k-1} \alpha^{iq^j}=a.
	$$

\end{proof}

We will describe the quasi-cyclic representation for the point-line incidences, again with block sizes of size $b$, by a $1 \times q^{k-1}$ matrix H$^{\mathrm{rep}}$ whose columns are indexed by
$$
	L_1=\{ a \in {\mathbb F}_{q^k} \ | \ \sum_{j=0}^{k-1} a^{q^j}=0\}.
$$

\begin{theorem} \label{agqcthm}
	The quasi-cyclic representation of the point-line incidences of $\mathrm{AG}(k,q)$ is given by a $1 \times q^{k-1}$ matrix $\mathrm{H}^{\mathrm{rep}}$, whose columns are indexed by elements of $L_1$ and where the $(1,a)$ entry $\mathrm{H}^{\mathrm{rep}}_{1,a}$
	of $\mathrm{H}^{\mathrm{rep}}$ is a subset of $\{0,\ldots,b-1\}$ which contains $i$ if and only if
	$$
		\alpha^{iq}-\alpha^i=a,
	$$
	and where $b=(q^k-1)/(q-1)$.
\end{theorem}

\begin{proof}
	As in the previous proof the points of $\mathrm{AG}(k,q)$ are the elements of ${\mathbb F}_q^k$ which we will consider as elements of $x \in {\mathbb F}_{q^k}$. Then the lines $\ell_{a,c}$ are defined by
	$$
		(cx)^q-cx=a,
	$$
	where $a \in L_1$ and  $c^b=1$. Since for all non-zero $\lambda \in {\mathbb F}_q$, the line defined by $(a,c)$ is the same as the hyperplane $(\lambda a, \lambda c)$, we can suppose $c^b=1$.

	For $\alpha$, a primitive $b$-th root of unity,
	\begin{equation} \label{affinelineeqn}
		x \in \ell_{a,\alpha c} \Leftrightarrow  (\alpha cx)^{q}-\alpha cx=a \Leftrightarrow
		\alpha x \in \ell_{a,c}.
	\end{equation}
	Thus, we can obtain an orbit of size $b$ by setting $c=x=1$ and taking the lines $\ell_{a,1}$ as representatives for each orbit, where $a \in L_1$.

	Thus, we index the columns of H$^{\mathrm{rep}}$  by the elements of $a\in L_1$ where the
	H$^{\mathrm{rep}}_{1,a}$
	of H$^{\mathrm{rep}}$ is a subset of $\{0,\ldots,b-1\}$ which contains $i$ if and only if
	$$
		\alpha^{iq}-\alpha^i=a.
	$$

\end{proof}

\section{The parameters of the classical GQ LDPC codes of length up to 400,000.} \label{sectionparam}

In the following tables we list the parameters of all LDPC codes which arise from classical generalized quadrangles of length up to 400,000. In the case of all quadrangles apart from $H(4,q^2)$ and its dual, a spread is removed so that the block size can be increased by a factor of $q$. As mentioned before, a quasi-cyclic generator matrix for the full code $C$ may not exist. In these cases, we list the rate of a sub-code $C'$ for which we have calculated a quasi-cyclic generator matrix. The check and generator matrices for these codes are publicly available~\cite{Ortega_QuasiCyclicGQs}.

\begin{center}
	\begin{table}[htbp!]
		\centering
		%\footnotesize
		\caption{The symplectic quadrangle $W(3,q) \setminus L_C$.} \label{tablewq}
		\begin{tabular}{|c|c|c|c|c|c|c|c|c|c|}
			\hline
			$q$ & length $n$ & block size $b$ & $\mathrm{H}^{\mathrm{rep}}$ & rank H & dim $C$ & rate $C$ & dim $C'$ & rate $C'$ & $\mathrm{P}^{\mathrm{rep}}$ \\ \hline
			5   & 156        & 26             & $5 \times 6$                & 91     & 65      & 0.41667  & 52       & 0.33333   & $2\times 4$                 \\
			9   & 820        & 82             & $9 \times 10$               & 451    & 369     & 0.45     & 328      & 0.4       & $4\times 6$                 \\
			11  & 1464       & 122            & $11 \times 12 $             & 793    & 671     & 0.45833  & 610      & 0.41667   & $5\times 7$                 \\
			19  & 7240       & 362            & $19 \times 20$              & 3801   & 3439    & 0.475    & 3258     & 0.45      & $9 \times 11$               \\
			23  & 12720      & 530            & $23 \times 24$              & 6625   & 6095    & 0.47917  & 5830     & 0.45833   & $11 \times 13$              \\
			25  & 16276      & 626            & $25 \times 26$              & 8451   & 7825    & 0.48077  & 7512     & 0.46154   & $12 \times 14$              \\
			31  & 30784      & 962            & $31 \times 32$              & 15873  & 14911   & 0.48438  & 14430    & 0.46875   & $15 \times 17$              \\
			41  & 70664      & 1682           & $41 \times 42$              & 36163  & 34481   & 0.51176  & 33640    & 0.47619   & $20 \times 22$              \\ \hline
		\end{tabular}
	\end{table}
\end{center}

\begin{center}
	\begin{table}[htbp!]
		\centering
		%\footnotesize
		\caption{The dual of the symplectic quadrangle $W(3,q) \setminus L_C$.} \label{tablewqdual}
		\begin{tabular}{|c|c|c|c|c|c|c|c|c|c|}
			\hline
			$q$ & length $n$ & block size $b$ & $\mathrm{H}^{\mathrm{rep}}$ & rank H & dim $C$ & rate $C$ & dim $C'$ & rate $C'$ & $\mathrm{P}^{\mathrm{rep}}$ \\ \hline

			5   & 130        & 26             & $6 \times 5$                & 91     & 39      & 0.3      & 26       & 0.2       & $1 \times 4$                \\
			9   & 738        & 82             & $10 \times 9$               & 451    & 287     & 0.38889  & 246      & 0.33333   & $3 \times 6$                \\
			11  & 1342       & 122            & $12 \times 11 $             & 793    & 549     & 0.40909  & 488      & 0.36364   & $4 \times 7$                \\
			19  & 6878       & 362            & $20 \times 19$              & 3801   & 3077    & 0.44737  & 2896     & 0.42105   & $8 \times 11$               \\
			23  & 12190      & 530            & $24 \times 23$              & 6625   & 5565    & 0.45652  & 5300     & 0.43478   & $10 \times 13$              \\
			25  & 15650      & 626            & $26 \times 25$              & 8451   & 7199    & 0.46     & 6886     & 0.44      & $11 \times 14$              \\
			31  & 29822      & 962            & $32 \times 31$              & 15873  & 13949   & 0.46774  & 13468    & 0.45161   & $14 \times 17$              \\
			41  & 68962      & 1682           & $42 \times 41$              & 36163  & 32799   & 0.52439  & 31958    & 0.46341   & $19 \times 22$              \\\hline
		\end{tabular}
	\end{table}
\end{center}

\begin{center}
	\begin{table}[htbp!]
		\centering
		%\footnotesize
		\caption{The dual of the elliptic quadric quadrangle $Q(5,q) \setminus L_C$.} \label{tableq5q}
		\begin{tabular}{|c|c|c|c|c|c|c|c|c|c|}
			\hline
			$q$ & length $n$ & block size $b$ & $\mathrm{H}^{\mathrm{rep}}$ & rank H & dim $C$ & rate $C$ & dim $C'$ & rate $C'$ & $\mathrm{G}^{\mathrm{rep}}$ \\ \hline
			3   & 112        & 28             & $9 \times 4$                & 91     & 21      & 0.1875   & 21       & 0.1875    & $1 \times 4$                \\
			5   & 756        & 126            & $25 \times 6$               & 651    & 105     & 0.13889  & 105      & 0.13889   & $1 \times 6$                \\
			7   & 2752       & 344            & $49 \times 8$               & 2451   & 301     & 0.10938  & 301      & 0.10938   & $1 \times 8$                \\
			9   & 7300       & 730            & $81\times 10$               & 6643   & 657     & 0.09     & 657      & 0.09      & $1 \times 10$               \\
			%11 &  15984 & 1332 & $121\times 12$ & 14763  &  1221 &  &   & $1 \times 12 $\\  
			13  & 30772      & 2198           & $169\times 14$              & 28731  & 2041    & 0.0663   & 2041     & 0.0663    & $1 \times 14$               \\  \hline
		\end{tabular}
	\end{table}
\end{center}

\begin{center}
	\begin{table}[htbp!]
		\centering
		%\footnotesize
		\caption{The elliptic quadric quadrangle $Q(5,q) \setminus L_C$.} \label{tableq5qdual}
		\begin{tabular}{|c|c|c|c|c|c|c|c|c|c|}
			\hline
			$q$ & length $n$ & block size $b$ & $\mathrm{H}^{\mathrm{rep}}$ & rank H & dim $C$ & rate $C$ & dim $C'$ & rate $C'$ & $\mathrm{P}^{\mathrm{rep}}$ \\ \hline
			3   & 252        & 28             & $4 \times 9$                & 91     & 161     & 0.63889  & 140      & 0.55555   & $5 \times 4$                \\
			5   & 3150       & 126            & $6 \times 25$               & 651    & 2499    & 0.79333  & 2394     & 0.76      & $19 \times 6$               \\
			7   & 16856      & 344            & $8 \times 49$               & 2451   & 14405   & 0.85459  & 14104    & 0.83673   & $41 \times 8$               \\
			9   & 59130      & 730            & $10 \times 81$              & 6643   & 52487   & 0.88766  & 51830    & 0.87654   & $71 \times 10$              \\
			11  & 161172     & 1332           & $12\times 121$              & 14763  & 146409  & 0.90840  & 145188   & 0.90083   & $109 \times 12 $            \\
			13  & 371462     & 2198           & $14\times 169$              & 28731  & 342731  & 0.92265  & 340690   & 0.91716   & $155 \times 14$             \\  \hline
		\end{tabular}
	\end{table}
\end{center}

\begin{center}
	\begin{table}[htbp!]
		\centering
		%\footnotesize
		\caption{The Hermitian quadrangle $H(4,q^2)$.} \label{tableh4q}
		\begin{tabular}{|c|c|c|c|c|c|c|c|c|c|}
			\hline
			$q^2$ & length $n$ & block size $b$ & $\mathrm{H}^{\mathrm{rep}}$ & rank H & dim $C$ & rate $C$ & dim $C'$ & rate $C'$ & $\mathrm{P}^{\mathrm{rep}}$ \\ \hline
			4     & 165        & 11             & $27 \times 15$              & 120    & 45      & 0.27273  & 44       & 0.26667   & $4 \times 11$               \\
			9     & 2440       & 61             & $112 \times 40$             & 1891   & 549     & 0.225    & 549      & 0.225     & $9 \times 32$               \\
			25    & 81276      & 521            & $756 \times 156$            & 68251  & 13025   & 0.16026  & 13025    & 0.16026   & $25 \times 31$              \\  \hline
		\end{tabular}
	\end{table}
\end{center}

\begin{center}
	\begin{table}[htbp!]
		\centering
		%\footnotesize
		\caption{The dual of the Hermitian quadrangle $H(4,q^2)$.} \label{tableh4qdual}
		\begin{tabular}{|c|c|c|c|c|c|c|c|c|c|}
			\hline
			$q^2$ & length $n$ & block size $b$ & $\mathrm{H}^{\mathrm{rep}}$ & rank H & dim $C$ & rate $C$ & dim $C'$ & rate $C'$ & $\mathrm{P}^{\mathrm{rep}}$ \\ \hline
			4     & 297        & 11             & $15 \times 27$              & 120    & 177     & 0.59596  & 176      & 0.59259   & $16 \times 11$              \\
			9     & 6832       & 61             & $40 \times 112$             & 1891   & 4941    & 0.72321  & 4941     & 0.72321   & $81\times 31$               \\
			25    & 393876     & 521            & $156 \times 756$            & 68251  & 325625  & 0.82672  & 325625   & 0.82672   & $625 \times 131$            \\  \hline
		\end{tabular}
	\end{table}
\end{center}

\section{Performance of classical GQ LDPC codes} \label{sectionperform}
In this section we consider the performance of some codes in the Binary Input AWGN channel. We plot the frame error rate against the signal-to-noise ratio, measured in decibels (dB), and compare this to the Polyanskiy-Poor-Verdú (PPV) bound~\cite{polyanskiy2010channel}  and DVB-S2 codes with a similar rate and length.
Note that the innovation with the codes discovered here is that very long GQ LDPC codes can be implemented.
All simulations were carried out using the AFF3CT toolbox~\cite{Cassagne2019a}, except the Q$(5, 13)$ code, which was generously tested by Dariush Divsalar using a field programmable gate array.

\begin{figure}[htbp]
	\centering
	\includegraphics[width=3.5in]{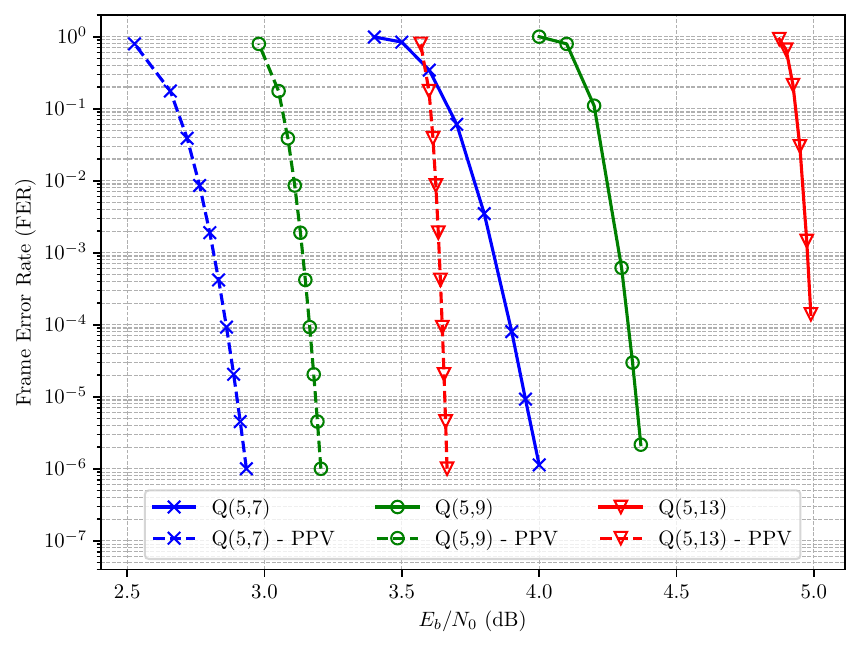}
	\caption{The performance of the LDPC codes from Q$(5,7)$, Q$(5,9)$, and Q$(5,13)$. The corresponding PPV bounds are plotted in dashed lines. Each code is decoded using 25 SPA iterations.}
	\label{fig:Qs25ite}
\end{figure}
\Cref{fig:Qs25ite} shows the performance curves for three example codes: Q$(5,7)$, Q$(5,9)$, and Q$(5,13)$. Each code is decoded using 25 SPA iterations. The figure also includes PPV bounds corresponding to their specific code lengths and rates. Notably, all the codes exhibit an approximate 1dB optimality gap, and no visible error floor.
The performance curves indicate that these very long codes would be suitable for applications such as memory devices.

\begin{figure}[htbp]
	\centering
	\includegraphics[width=3.5in]{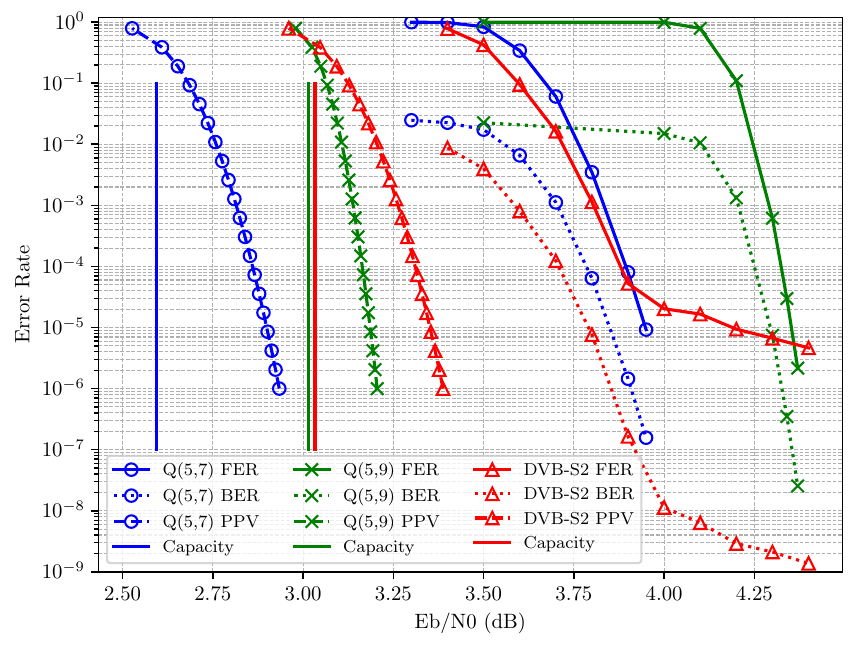}
	\caption{The performance of the LDPC codes from Q$(5,7)$, Q$(5,9)$, and (16200, 14400) DVB-S2. Each code is decoded using 50 SPA iterations.}
	\label{fig:perf}
\end{figure}
\Cref{fig:perf} compares the performance of the Q$(5,7)$ code, the Q$(5,9)$ code, and the commercially used (16200, 14400) DVB-S2 code.
The code from Q$(5,7)$ has dimensions (16856, 14405), which are very similar to the chosen DVB-S2 code, despite having a slightly lower rate.
For a closer rate code, we also plot Q$(5,9)$, which is a (59130, 52487) code.
We observe that both GQ codes outperform DVB-S2 in the operating range above 4.5dB.
Also, both codes perform around 1.5 dB from capacity.

\begin{figure}[htbp]
	\centering
	\includegraphics[width=3.9in]{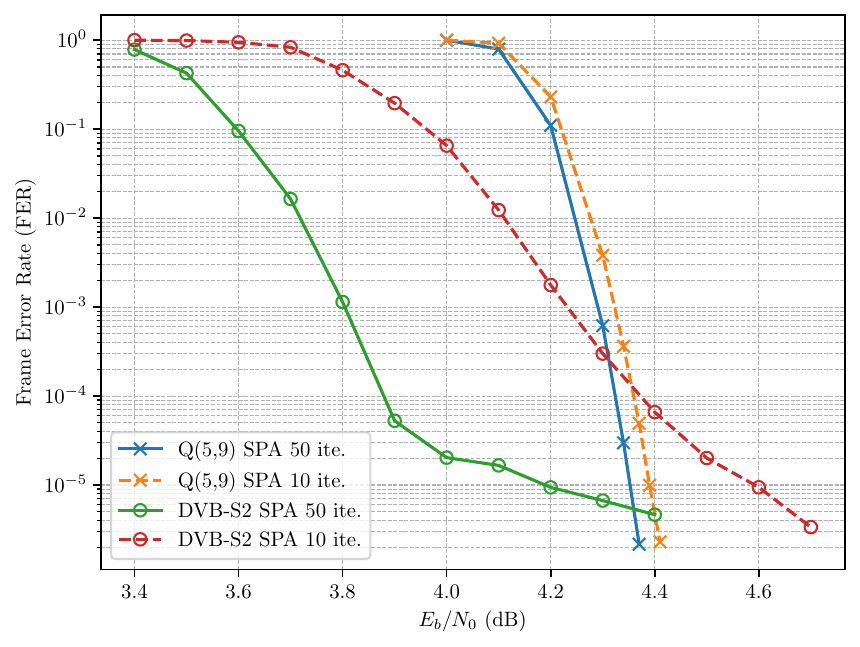}
	\caption{Q$(5,9)$ code performance compared with a (16200, 14400) DVB-S2 code of the same rate, decoded with different number of SPA iterations.}
	\label{fig:perf2}
\end{figure}
\begin{figure}[htbp]
	\centering
	\includegraphics[width=3.9in]{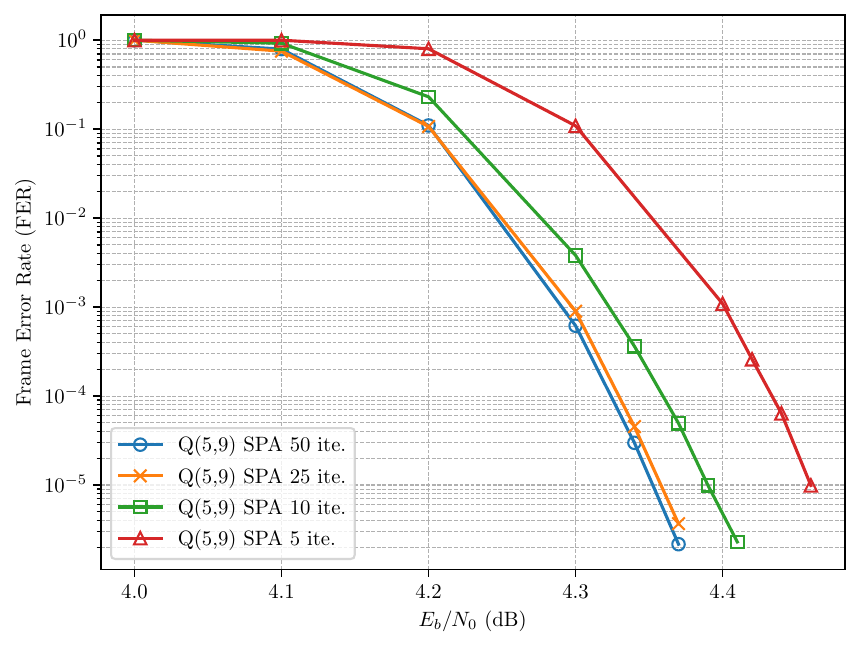}
	\caption{Q$(5,9)$ performance curves for different number of SPA iterations.}
	\label{fig:ites}
\end{figure}
As mentioned in the introduction, the other advantage of using GQ LDPC codes is that the decoding algorithms run fast for such codes, see~\cref{fig:perf2}.
Notably, the performance gap between 10 and 50 SPA iterations is much smaller for our codes than for DVB-S2 codes.
\Cref{fig:ites} shows how lowering the number of SPA iterations lowers the decoding performance.

\section{Further work} \label{sectionfurther}

\begin{enumerate}
	\item
	      We are particularly keen that these LDPC codes are implemented. We have presented evidence that these very long codes perform well with low-complexity decoding algorithms.
	\item
	      This quasi-cyclic representation of $W(3,q)$ appeared for the first time in \cite{Ball2004}. There it was used to prove that any spread of $W(3,q)$ share $1$ modulo $p$ lines with a regular spread of $W(3,q)$, where $p$ is the prime such that $q=p^h$. This result was subsequently used in \cite{BGS2006} to prove that if $q$ is prime then all spreads of $W(3,q)$ are regular.
	      It is probable that one can obtain more results regarding spreads and ovoids of classical generalized quadrangles using the quasi-cyclic representations.
	\item
	      It may be a useful exercise to compute directly the quasi-cyclic representations for the dual quadrangles. Although this will result in the same check matrices, one may see that the lines split into two or more classes, as in the case of the three quadrangles we have calculated here. Again, this may help prove results concerning spreads and ovoids.
	\item
	      It should be possible to calculate the quasi-cyclic representation of the classical generalized octagons.
	\item
	      It should be possible to calculate the finite field extension representations of all classical polar spaces. One can expect that removing small substructures, as in the case of the symplectic and elliptic quadrangles, will reveal quasi-cyclic structures with large block sizes.
	\item
	      We have included the quasi-cyclic representation for the point-hyperplane and the point-line incidence matrices of both the projective and affine spaces. It should be possible to calculate the quasi-cyclic representation of more general incidences. One may find interesting quasi-cyclic representations for $s$-dimensional subspaces contained in $r$-dimensional subspaces, for certain $r$ and $s$ by removing a small set of subspaces. The codes $\mathcal C_{r,s}$, over ${\mathbb F}_p$, whose generator matrices are the codes obtain from these containments have a rich history. The dimensions of these codes was calculated by Hamada \cite{Hamada68} and for the affine space \cite{Hamada73}. For recent results concerning small weight codewords in these codes, see \cite{AD2021} and \cite{AD2024}.

\end{enumerate}

\section*{Acknowledgment}
We thank Tabriz Popatia for his careful reading of the manuscript and Dariush Divsalar for testing the Q$(5,13)$ code.

%---------------------------- Bibliography -------------------------------

\bibliographystyle{IEEEtran}
\bibliography{references}

\begin{thebibliography}{10}
\providecommand{\url}[1]{#1}
\csname url@samestyle\endcsname
\providecommand{\newblock}{\relax}
\providecommand{\bibinfo}[2]{#2}
\providecommand{\BIBentrySTDinterwordspacing}{\spaceskip=0pt\relax}
\providecommand{\BIBentryALTinterwordstretchfactor}{4}
\providecommand{\BIBentryALTinterwordspacing}{\spaceskip=\fontdimen2\font plus
\BIBentryALTinterwordstretchfactor\fontdimen3\font minus
  \fontdimen4\font\relax}
\providecommand{\BIBforeignlanguage}[2]{{%
\expandafter\ifx\csname l@#1\endcsname\relax
\typeout{** WARNING: IEEEtran.bst: No hyphenation pattern has been}%
\typeout{** loaded for the language `#1'. Using the pattern for}%
\typeout{** the default language instead.}%
\else
\language=\csname l@#1\endcsname
\fi
#2}}
\providecommand{\BIBdecl}{\relax}
\BIBdecl

\bibitem{kou2001low}
Y.~Kou, S.~Lin, and M.~P. Fossorier, ``Low-density parity-check codes based on
  finite geometries: a rediscovery and new results,'' \emph{IEEE Transactions
  on Information theory}, vol.~47, no.~7, pp. 2711--2736, 2001.

\bibitem{LP2005}
Z.~Liu and D.~Pados, ``{LDPC} codes from generalized polygons,'' \emph{IEEE
  Transactions on Information Theory}, vol.~51, pp. 3890--3898, 2005.

\bibitem{FH1964}
W.~Feit and G.~Higman, ``The nonexistence of certain generalized polygons,''
  \emph{Journal of Algebra}, vol.~1, pp. 114--131, 1964.

\bibitem{RL2009}
W.~E. Ryan and S.~Lin, \emph{Channel codes: Classical and Modern}.\hskip 1em
  plus 0.5em minus 0.4em\relax Cambridge University Press, 2009.

\bibitem{LCZL2006}
Z.~Li, L.~Chen, L.~Zeng, S.~Lin, and W.~H. Fong, ``Efficient encoding of
  quasi-cyclic low-density parity-check codes,'' \emph{IEEE Transactions on
  Information Theory}, vol.~54, pp. 71--81, 2006.

\bibitem{Ortega_QuasiCyclicGQs}
\BIBentryALTinterwordspacing
T.~Ortega. (2024) Quasi cyclic representation of classical generalized
  quadrangles. GitHub repository. [Online]. Available:
  \url{https://github.com/TomasOrtega/QuasiCyclicGQs}
\BIBentrySTDinterwordspacing

\bibitem{polyanskiy2010channel}
Y.~Polyanskiy, H.~V. Poor, and S.~Verd{\'u}, ``Channel coding rate in the
  finite blocklength regime,'' \emph{IEEE Transactions on Information Theory},
  vol.~56, no.~5, pp. 2307--2359, 2010.

\bibitem{Cassagne2019a}
\BIBentryALTinterwordspacing
A.~Cassagne \emph{et~al.}, ``{AFF3CT}: A fast forward error correction
  toolbox!'' \emph{Elsevier SoftwareX}, vol.~10, p. 100345, Oct. 2019.
  [Online]. Available:
  \url{http://www.sciencedirect.com/science/article/pii/S2352711019300457}
\BIBentrySTDinterwordspacing

\bibitem{Ball2004}
S.~Ball, ``On ovoids of {O(5,q)},'' \emph{Advances in Geometry}, vol.~4, pp.
  1--7, 2004.

\bibitem{BGS2006}
S.~Ball, P.~Govaerts, and L.~Storme, ``On ovoids of parabolic quadrics,''
  \emph{Designs, Codes and Cryptography}, vol.~38, pp. 131--145, 2006.

\bibitem{Hamada68}
N.~Hamada, ``The rank of the incidence matrix of points and $d$-flats in finite
  geometries,'' \emph{Journal of Science of the Hiroshima University, Series AI
  (Mathematics)}, vol.~32, pp. 381--396, 1968.

\bibitem{Hamada73}
------, ``On the $p$-rank of the incidence matrix of a balanced or partially
  balanced incomplete block design and its application to error-correcting
  codes,'' \emph{Hiroshima Mathematics Journal}, vol.~3, pp. 153--226, 1973.

\bibitem{AD2021}
S.~Adriaensen and L.~Denaux, ``Small weight codewords of projective geometric
  codes,'' \emph{Journal of Combinatorial Theory, Series A}, vol. 180, pp.
  paper no. 105\,395, 34, 2021.

\bibitem{AD2024}
------, ``Small weight codewords of projective geometric codes ii,''
  \emph{Designs, Codes and Cryptography}, 2024, to appear.
  \url{arxiv:2309.00490}.

\end{thebibliography}

\end{document}